\documentclass[aps,prf,floatfix,superscriptaddress,reprint]{revtex4} 
\usepackage[utf8]{inputenc}
\usepackage{graphicx}
\usepackage{amsmath}
\usepackage{siunitx}
\usepackage[usenames,dvipsnames]{xcolor}
\usepackage{commath}
\usepackage{placeins}
\definecolor{MyBlue}{rgb}{0,0.3,0.6}
\usepackage[colorlinks=true,linkcolor=MyBlue,plainpages=false,citecolor=MyBlue,urlcolor=MyBlue]{hyperref}

\usepackage{multirow}
\usepackage{mathtools}
\makeatletter
\let\oldabs\abs
\def\abs{\@ifstar{\oldabs}{\oldabs*}}

\newcommand{\Nus}{\text{Nu}}
\newcommand{\Pra}{\text{Pr}}
\newcommand{\Ray}{\text{Ra}}
\newcommand{\Rey}{\text{Re}}

\begin{document}
\title{Heat transfer enhancement in Rayleigh-B\'enard convection\\using a single passive barrier}

\author{Shuang Liu}
\email{wsls52690@gmail.com}
\affiliation{Center for Combustion Energy, Key Laboratory for Thermal Science
and Power Engineering of Ministry of Education, Department of Energy and Power
Engineering, Tsinghua University, Beijing 100084, China}

\author{Sander~G.~Huisman}
\email{s.g.huisman@gmail.com}
\affiliation{Physics of Fluids Group, Max Planck UT Center for Complex Fluid Dynamics,\break
MESA+ Institute and J.M. Burgers Centre for Fluid Dynamics, University of Twente, 7500 AE Enschede, The Netherlands}

\date{\today}

\begin{abstract}
In this numerical study on Rayleigh-B\'enard convection we seek to improve the heat transfer by passive means. To this end we introduce a single tilted conductive barrier centered in an aspect ratio one cell, breaking the symmetry of the geometry and to channel the ascending hot and descending cold plumes. We study the global and local heat transfer and the flow organization for Rayleigh numbers $10^5 \leq \Ray \leq 10^9$ for a fixed Prandtl number of $\text{\Pra=4.3}$. We find that the global heat transfer can be enhanced up to $18\%$, and locally around $800\%$. The averaged Reynolds number is always decreased when a barrier is introduced, even for those cases where the global heat transfer is increased. We map the entire parameter space spanned by the orientation and the size of a single barrier for $\Ray=10^8$.
\end{abstract}
\maketitle

\section{Introduction}\label{sec:introduction}
\begin{figure}[htb]
\centering
\includegraphics[width=0.45\textwidth]{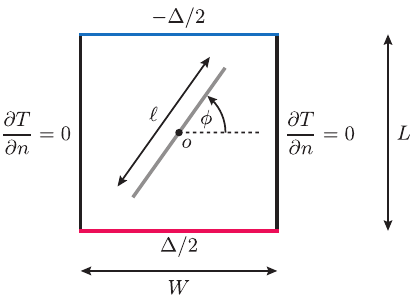}
\vspace{-2 mm}
\caption{Rayleigh-B\'enard geometry with a single infinitesimally thin barrier (gray). The center of the cell is indicated by a black disk and denoted $o$. The barrier is centered around $o$, has a length $\ell$, and is oriented with an angle $\phi$ with respect to the horizontal (dashed line). The convection cell has unit width $W$ and therefore the aspect ratio $\Gamma=W/L=1$.
}
\label{fig:setup}
\end{figure}

Rayleigh-B\'enard (RB) convection consists of a fluid cell heated from the bottom and cooled from the top. The fluid is set into motion due to gravity acting on the density differences that arise in the heated fluid. This motion greatly enhances the thermal transport. This simplified geometry is paradigmatic for natural convection, taking the most basic elements of the problem of turbulent heat transfer while maintaining a large variety of observable phenomena. The system has applications in geophysics including mantle convection and weather forecasting, astrophysics, and industry, and has been widely studied over the last decades \cite{ahlers2009,lohsearfm2010,xiataaml2013}. The driving of RB convection, assuming the Boussinesq approximation, is given by the dimensionless Rayleigh number $\Ray=g\beta \Delta L^3/(\nu\kappa)$ where $g$ is the gravitational acceleration, $L$ the height of the convection cell, $\beta$ the thermal expansion coefficient, $\Delta$ the temperature difference between the top and bottom of the cell (see Fig.~\ref{fig:setup}), $\nu$ the kinematic viscosity of the fluid, and $\kappa$ the thermal diffusivity of the fluid. The dynamics also depend geometrically on the aspect ratio of the cell $\Gamma=W/L$, where $W$ is the width of the convection cell \cite{van2011connecting}, and furthermore they also depend on the material properties of the fluid by the Prandtl number $\Pra=\nu/\kappa$. When the temperature difference $\Delta$ is positive, heat will flow from the bottom to the top boundary of the convection cell. Up to a critical Rayleigh number the heat is transferred by pure conduction whereas for large $\Ray$ the fluid is set into motion and convection will quickly become the dominating mode of heat transfer in the geometry. The heat transfer is characterised by the dimensionless Nusselt number:
\begin{align}
\Nus=\frac{Q}{k\Delta L^{-1}}=-\left\langle \frac{\partial T}{\partial z} \right\rangle_{W,t},
\end{align}
where $Q$ denotes the dimensional total heat flux, $k$ is the thermal conductivity, $T$ is the dimensionless temperature normalized by the temperature difference $\Delta$ and shifted such that $0 \leq T \leq 1$, $z$ is the vertical coordinate normalized by the height of the system $L$, and $\langle\hdots\rangle_{W,t}$ denotes the average over the horizontal plate and over time. $\Nus$ compares the total heat flux with the heat flux for the pure conductive case. In general, even for such a simple geometry, the heat transfer is a complicated function $f$:  $\Nus=f(\Ray,\Pra,\Gamma)$ \cite{GL2000}.

Often the goal is to enhance the heat transfer. This can be done using a variety of strategies such as adding bubbles \cite{lakkaraju2013,gvozdic2018,lohse2018bubble,wang2018bubble} or multi-component fluids with phase-change \cite{narezo2016a,narezo2016b,hoefnagels2017a,wang2019self,wang2020}, adding roughness \cite{wagner2015,zhu2017roughness,jiang2018,zhang2018surface,zhu2019}, vibration \cite{wang2020vibration,yang2020periodically}, changing the velocity boundary conditions \cite{vanderp2014}, and coherent structure manipulation by an additional stabilization due to confinement, rotation, double-diffusive effect or an external magnetic field \cite{chong2017confined,huang2013confinement,chong2015condensation,stevens2009transitions,zhong2009prandtl,stevens2013b,lim2019quasistatic,yang2020rotation,yang2020double}. 

Detaching from the thermal boundary layers, the thermal plumes play an important role to the sustained heat transfer. Under the action of buoyancy force they mainly produce a directional motion with the hot and cold plumes moving upwards and downwards, respectively, driving a large scale circulation (LSC). Sometimes, however, counter-gradient motions of thermal plumes have been found, with ascending cold plumes and descending hot plumes. These misoriented motions of plumes are associated with the bulk dynamics and the competition between the corner rolls and LSC, and they can lead to strongly negative local heat flux \cite{shang2003measured,huang2013counter}.
One way to control the plume dynamics is by adding barriers or obstacles, for which one might---naively---guess that adding them to the geometry probably decreases the flow and therefore the thermal transport, though the opposite can be true \cite{bao2015,liu2020}.

Another aspect of the misoriented dynamics of convective flow is the reversal of the LSC. Flow reversal in RB convection has received persistent interest, since it may provide important insights to many natural phenomena, such as the reversal processes of geomagnetic field and atmosphere \cite{glatzmaier1999role,van2000statistics}. With the complex three-dimensional (3D) flow dynamics suppressed \cite{funfschilling2004plume,brown2005reorientation,xi2009origin,xi2016higher}, flow reversal in two-dimensional (2D) and quasi-2D systems has received much attention recently, where the complex interaction between the LSC and corner rolls plays an important role to reversals \cite{sugiyama2010flow,chandra2011dynamics,chandra2013flow,verma2015flow,huang2015comparative,ni2015reversals,mannattil2017applicability,xia2016flow,podvin2017precursor,wang2018flow,chen2019emergence,chen2020reduced}.
While most of previous studies considered idealized symmetric systems, it is desirable to examine how symmetry-breaking effects influence flow reversals, since asymmetry is omnipresent in real world \cite{huang2015comparative,xia2016flow,wang2018flow}.
In many circumstances, adding asymmetry to the system can orient the LSC. This has been done before by inclining the apparatus \cite{shishkina2016,zwirner2018,wang2018multiple,jiang2019robustness}, and is also commonly done in experiments (though with much smaller inclinations) to fix the LSC in place for visualization and measurement purposes.

In the current work we explore the possibility of increasing the thermal transport and controlling flow dynamics via passive means. We study RB convection in a standard geometry of aspect ratio $\Gamma=W/L=1$. A single passive angled flow barrier is included inside the convection cell to stabilize the LSC, to guide the plume motions, and to enhance the heat transfer, see Fig.~\ref{fig:setup}.

The remainder of the paper is organized as follows. The numerical details are given in \S\ref{sec:numerical}. The main results are presented and discussed in \S\ref{sec:results}, focusing on the effects of barrier on the heat transfer, flow organization, and flow reversals. In \S\ref{sec:summary} the main findings are summarized. Finally, two appendices are provided to give more details on the grid resolution and the influence of the barrier on the flow structure.

\section{Numerical details}\label{sec:numerical}
As shown in Fig.~\ref{fig:setup}, we consider the RB convection in a cell of unit aspect ratio, where a straight, solid conductive barrier with infinitesimal thickness and length $\ell$ is added, with the barrier center collapsing on the cell center $o$.
The fluid flow and heat transport are described by the dimensionless incompressible Navier--Stokes equations, the convection–diffusion equation, and the continuity equation within the Boussinesq approximation:
\begin{eqnarray}
\begin{aligned}
\frac{\partial \vec{u}}{\partial t}+\vec{u} \cdot \vec{\nabla} \vec{u} &= -\vec{\nabla} p+\sqrt{\frac{\Pra}{\Ray}} \vec{\nabla}^{2} \vec{u}+T \vec{e}_{z}, \\
\frac{\partial T}{\partial t}+ \vec{u}\cdot\vec{\nabla} T &= \frac{1}{\sqrt{\Pra\,\Ray}} \vec{\nabla}^{2} T, \\
\vec{\nabla} \cdot \vec{u} &= 0,
\end{aligned}
\label{governing_equations}
\end{eqnarray}
where $\vec{u}=(u,w)$ is the velocity vector in the $(x,z)$-plane, $p$ is the fluid pressure, and $T$ is the temperature. $\vec{e}_z$ is the unit vector along the vertical $z$-direction.
The two dimensionless numbers in the governing equations (\ref{governing_equations}) are the Rayleigh number $\Ray$ and Prandtl number $\Pra$, which are given in \S\ref{sec:introduction}. For the non-dimensionalization, the cell height $L$, the temperature difference $\Delta$ between the top and bottom plates, and the free-fall velocity $\sqrt{g \beta \Delta L}$ are used as the characteristic quantities for length, temperature, and velocity, respectively.

The cell is assumed to have unit width ($W=L=1$) and the barrier configuration is quantified by two geometric parameters, namely, the barrier length $\ell$ and the angle $\phi$ the barrier makes with respect to the horizontal axis.
There is a maximum barrier length $\ell_m$, beyond which the convection cell is divided into two sub-regions without mass transfer between each other. The maximum length $\ell_m$ for complete blockage is dependent on the barrier angle $\phi$. For our square unit cell we have $\ell_m=\min\!\!\left[\abs*{\sec(\phi)},\abs*{\csc(\phi)}\right]$. 

No-slip and no-penetration boundary conditions are imposed at all solid boundaries, including the barrier surfaces. The horizontal top and bottom plates are isothermal and the sidewalls are thermally insulated. The fluid and the barrier are assumed to have the same thermal properties. Practically, no thermal boundary conditions are explicitly enforced for the thin barrier, and the temperature equation is solved in the entire domain.

The governing equations (\ref{governing_equations}) were solved using a second-order finite-difference method implemented in the open-source AFiD code \cite{verzicco1996,van2015}. 
Time integration is performed using the fractional-step third-order scheme for the explicit terms and the Crank--Nicolson scheme for the implicit terms.
The grid is stretched along the vertical direction so that finer resolution is used near the horizontal plates to resolve the boundary layers.
The barrier is taken into account using the immersed boundary method \cite{fadlun2000combined,zhu2017roughness,zhu2019}, which is versatile for the treatment of complex geometries, while the fluid motion can be conveniently solved based on a Cartesian grid.

Adequate grid resolutions are used to resolve the boundary layers and the bulk flows, satisfying the resolution requirement for the direct numerical simulations of thermal convection \cite{shishkina2010boundary,stevens2010radial}. A finer resolution is employed when $\ell$ is close to $\ell_m$ such that the barrier end is close to the wall with a small gap.
The flow in the gap is resolved with at least 10 grid points. We note that for certain parameters $(\ell,\phi)$ the thermal boundary layers on the horizontal plates can be highly non-uniform with the local boundary layer thickness near the barrier end much smaller than the averaged thickness, as shown in Fig.~\ref{local_flux}$(a)$.
The grid is chosen such that the whole thermal boundary layer is well resolved, which agrees with the recommendations in Ref.~\cite{shishkina2010boundary}. For the highest $\Ray$ ($\Ray=10^9$) 2250 grid points are used in the vertical direction for $\ell=1.35$ and $\phi=\pi/4$, and the thermal boundary layers are everywhere resolved with at least 11 grid points.
Simulation parameters including the grid resolutions are presented in Table \ref{tab:parameters}. More details of grid resolution and a comparison with the local Kolmogorov length scale $\eta$ are given in Appendix \ref{appA}.
To guarantee that the resolution used is sufficient, grid independence studies were performed at selected parameters, and the numerical details are also provided in Appendix \ref{appA}.
The statistics of the global response parameters are based on data over sufficiently long times and the difference of Nusselt numbers based on the first and second halves of the simulations after the initial transients is within 1\%.

\begin{table}[htpb!]
	\centering
	\renewcommand\arraystretch{1.2}
	\setlength{\tabcolsep}{3mm}
	\begin{tabular}{ccccc}
		\hline
	 	Type  & $\Ray$        &	$\ell$       & $\phi$      & $N_z$      \\ \hline
		2D    & $[10^5,10^9]$ & $[0,\ell_m]$ & $\pi/4$     & 180--2250  \\
		2D    & $10^8$        & $[0,\ell_m]$ & $[0,\pi/2]$ & 280--2000  \\
		3D    & $10^8$        & $[0,\ell_m]$ & $\pi/4$     & 300--640    \\ \hline
	\end{tabular}
	\caption{\label{tab:parameters} Numerical parameters for the simulations. The columns from left to right indicate the dimensionality of the simulation, the Rayleigh number $\Ray$, the barrier length $\ell$, the barrier angle $\phi$, and the number of grid points $N_z$ in the vertical direction. For the 3D problem, the normalized spanwise depth is $1/4$.}
\end{table}

In this study the effects of a passive barrier on the heat transfer and flow organization are investigated for varying Rayleigh numbers $\Ray$ and barrier geometric properties $(\ell,\phi)$, at a fixed Prandtl number $\Pra=4.3$. The barrier angle $\phi$ takes values from 0 to $\pi/2$. For given $\phi$, the barrier length is varied from $0$ to $\ell_m$, as shown in Table \ref{tab:parameters}.

We mainly consider the 2D problem, since the 2D simulations are much less computationaly demanding, allowing us to explore a larger parameter space.
According to Ref.~\cite{van2013}, for a large range of $\Ray$ and large $\Pra>1$, the $\Nus$ versus $\Ray$ scalings in 2D and 3D convection are comparable up to constant prefactors, which justifies our 2D simulations at $\Pra=4.3$ for the traditional RB convection without barrier. A set of 3D simulations were also performed for comparison.

\section{Results and discussion}\label{sec:results}
\subsection{Heat transfer}\label{sec:heat_transfer}
\begin{figure}[htpb!]
	\centering
	\includegraphics[width=0.45\linewidth]{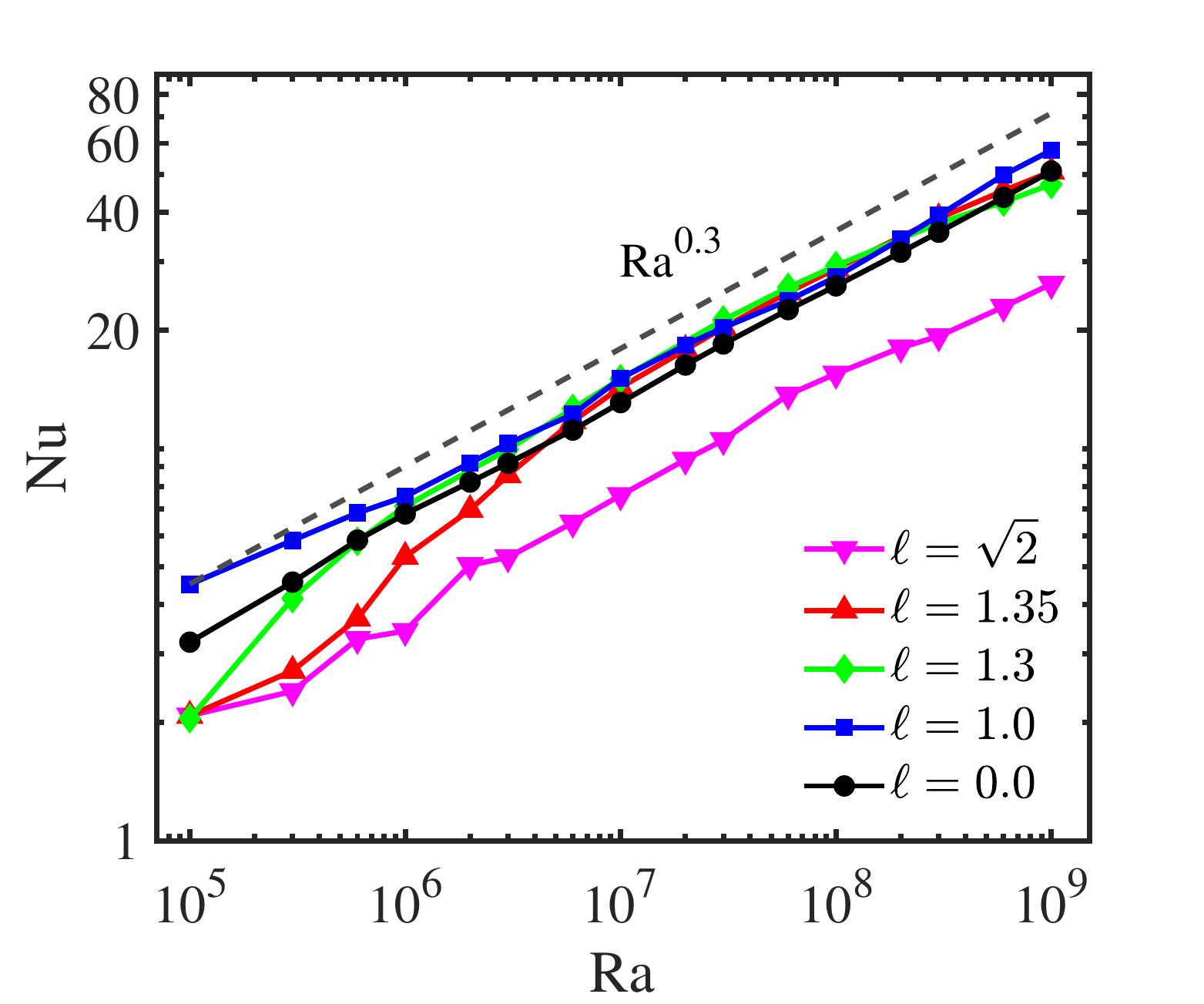}
	\put(-228,170){$(a)$}
	\includegraphics[width=0.45\linewidth]{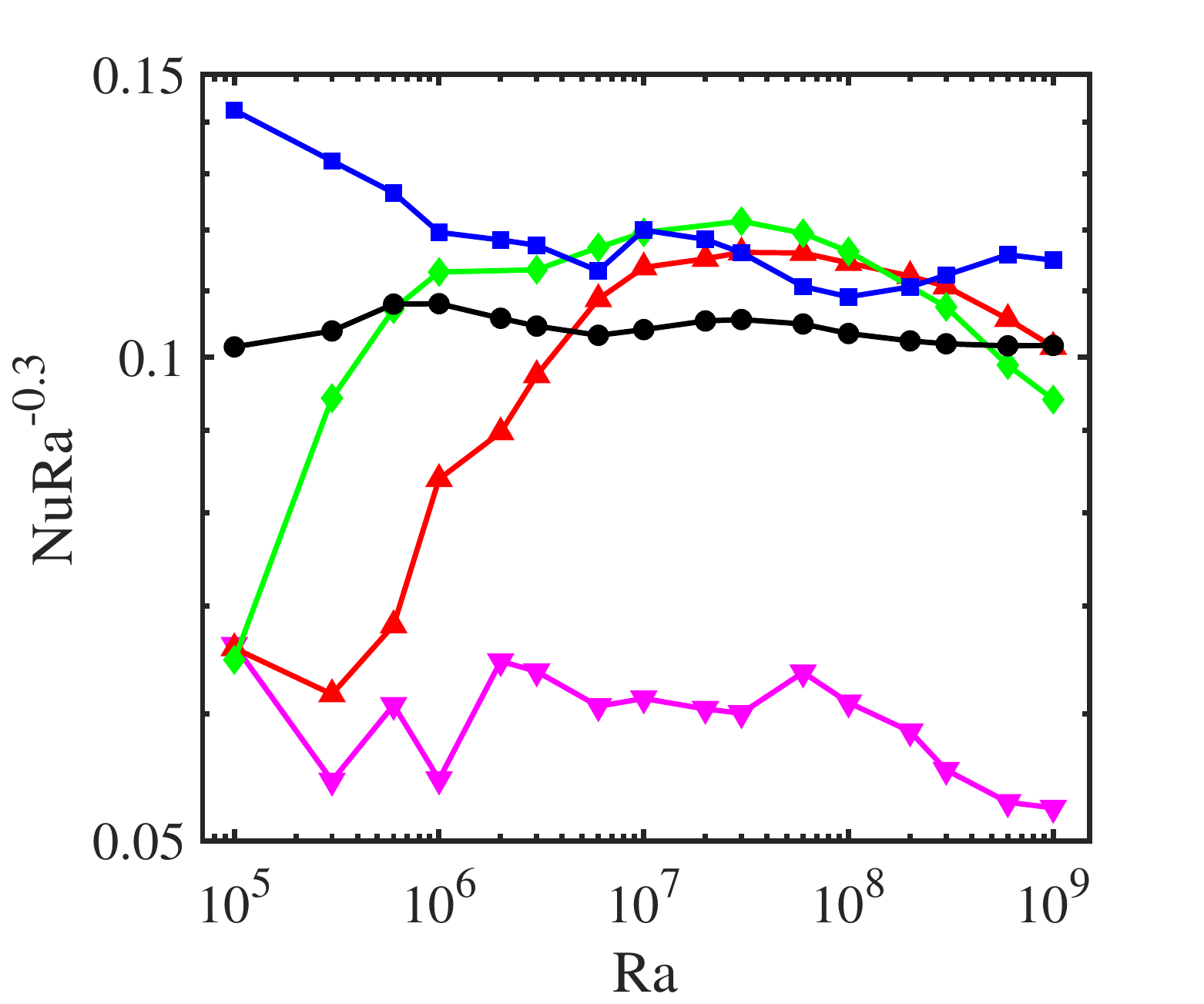}
	\put(-228,170){$(b)$}
	\vspace{-3 mm}
	\caption{\label{nu_ra_scaling} $(a)$ $\Nus$ as a function of $\Ray$ for different barrier lengths $\ell$ at a fixed angle $\phi=\pi/4$. For reference, a $\Nus \propto \Ray^{0.3}$ scaling is included as a dashed grey line. $(b)$ The compensated plot of $(a)$.}
\end{figure}

We first examine the influence of the barrier on the global heat transfer. 
Fig.~\ref{nu_ra_scaling}$(a)$ shows the dependence of $\Nus$ on $\Ray$ for various values of $\ell$ at a fixed barrier angle $\phi=\pi/4$. For $\phi=\pi/4$, the maximum barrier length $\ell_m=\sqrt{2}$.
In traditional RB convection with $\ell=0$, the increase of $\Nus$ with $\Ray$ follows an effective power law $\Nus \propto \Ray^{0.3}$ in the current $\Ray$ regime, consistent with previous results in the literature \cite{van2013,huang2013counter,zhang2017statistics,zhang2019efficient}. We provide the compensated plot showing $\Nus \Ray^{-0.3}$ in Fig.~\ref{nu_ra_scaling}$(b)$ revealing the details of the data.
When a barrier is introduced, the modification of $\Nus$ is dependent on both $\Ray$ and $\ell$. 
Interestingly, the addition of a flow barrier (a restriction) can lead to heat transfer enhancement. For $\ell=1.0$ the absolute values of $\Nus$ are uniformly larger than those of $\ell=0$ for the parameters examined (see black and blue data in Fig.~\ref{nu_ra_scaling}$(b)$), while the scaling behavior of $\Nus$ with $\Ray$ is similar to that of the traditional RB convection, suggesting that the heat transfer is still governed mainly by the thermal boundary layers \cite{ahlers2009}.
In the completely blocking configuration with $\ell=\ell_m$, the convection cell is divided into two sub-regions of a triangular geometry. Between the two sub-regions heat is solely transferred via thermal conduction through the barrier. On the horizontal plates of the top and bottom sub-regions the temperature is fixed at 0 and 1, respectively. Based on symmetry, the temperature on the inclined boundary (the barrier surface) is assumed to be 0.5 for simplification. 
Each triangular sub-system may be comparable to a traditional RB system with an effective temperature drop equal to $0.5$ and an effective cell height less than $L$ ($\approx 0.6L$), resulting in a smaller effective $\Ray$ (roughly 10 times smaller). Thus, it is expected that for $\phi=\pi/4$ the heat transfer is significantly reduced in the completely blocking configuration with $\ell=\ell_m=\sqrt{2}$, as observed in Fig.~\ref{nu_ra_scaling}.

When $\ell$ is close to $\ell_m$, the top and bottom sub-regions are connected via two small gaps between the barrier ends and the boundaries of the convection cell.
It is found that the dependence of $\Nus$ on $\Ray$ exhibits two different regimes, see the curves for $\ell=1.3$ and $\ell=1.35$ in Fig.~\ref{nu_ra_scaling}. When $\Ray$ is low, the heat transfer is reduced compared with that in traditional RB convection. The reduced heat transfer is attributed to the small gap and large coherence length scale of the flow at small $\Ray$. The convection between the two sub-regions is restricted by the small gap, and the system is close to the completely blocking configuration when $\ell$ is close to $\ell_m$. When $\Ray$ is low enough, the small gap is deeply immersed in the thermal boundary layer, and has negligible influence on the heat transfer, as shown in Fig.~\ref{nu_ra_scaling} that the Nusselt numbers for the three largest $\ell$ collapse for $\Ray=10^5$.
When $\Ray$ is high enough, the heat transfer is mildly affected by the barrier, and $\Nus$ follows a similar effective power law as that in the traditional RB convection. Due to the smaller coherence length scale of the flow at higher $\Ray$, a small gap becomes less restrictive.
The critical $\Ray$ separating the two regimes is dependent on $\ell$. When $\ell$ is closer to $\ell_m$, the gap becomes smaller, and correspondingly the critical $\Ray$ becomes larger.
Note that in the presence of a conductive barrier, the thermal conduction state is still a solution of the system. The convection onset will be delayed to higher $\Ray$ in the presence of a barrier due to the enhanced drag on the convection.

\begin{figure} [htpb!]
	\centering
	\includegraphics[width=0.45\linewidth]{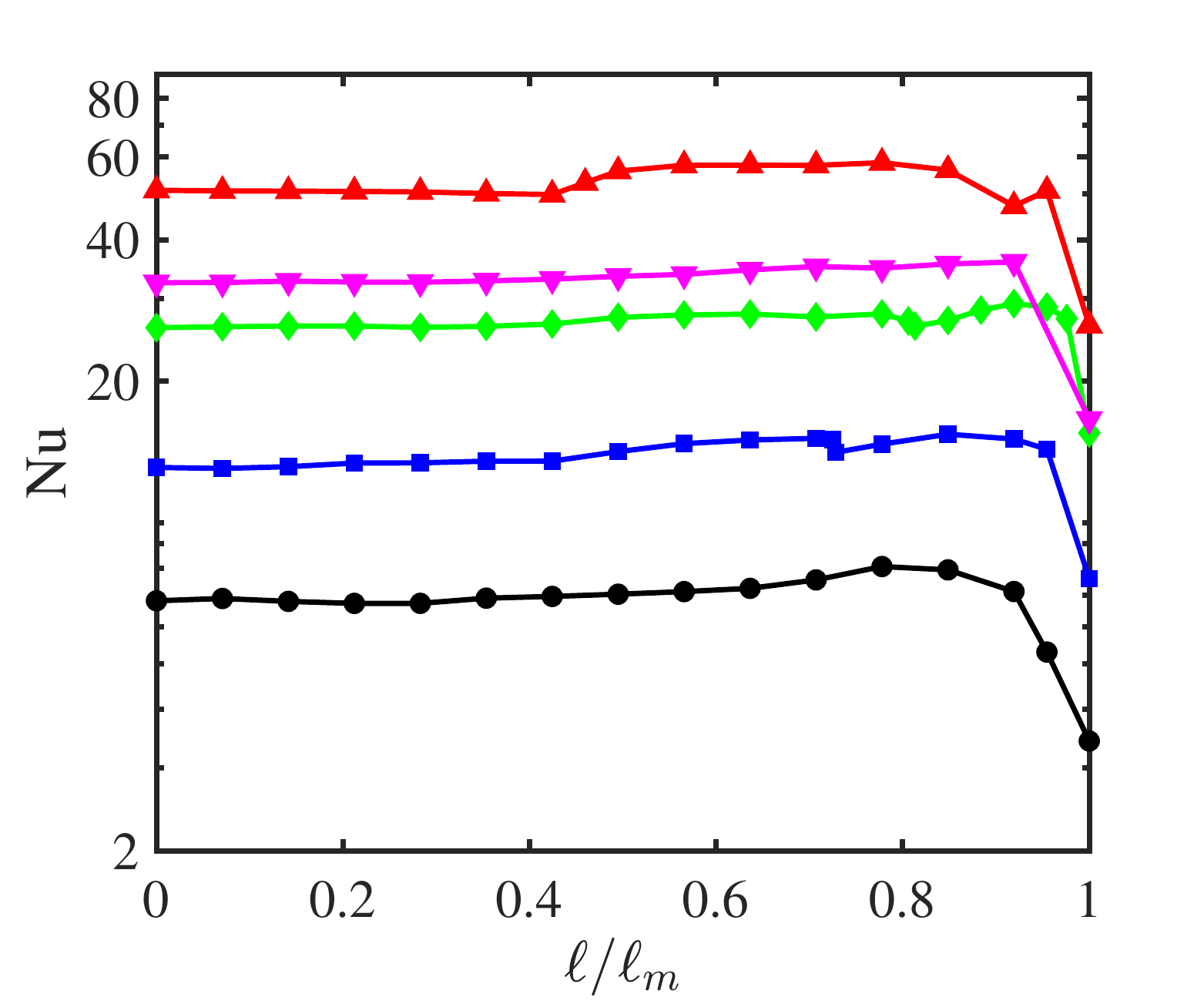}
	\put(-228,170){$(a)$}
	\includegraphics[width=0.45\linewidth]{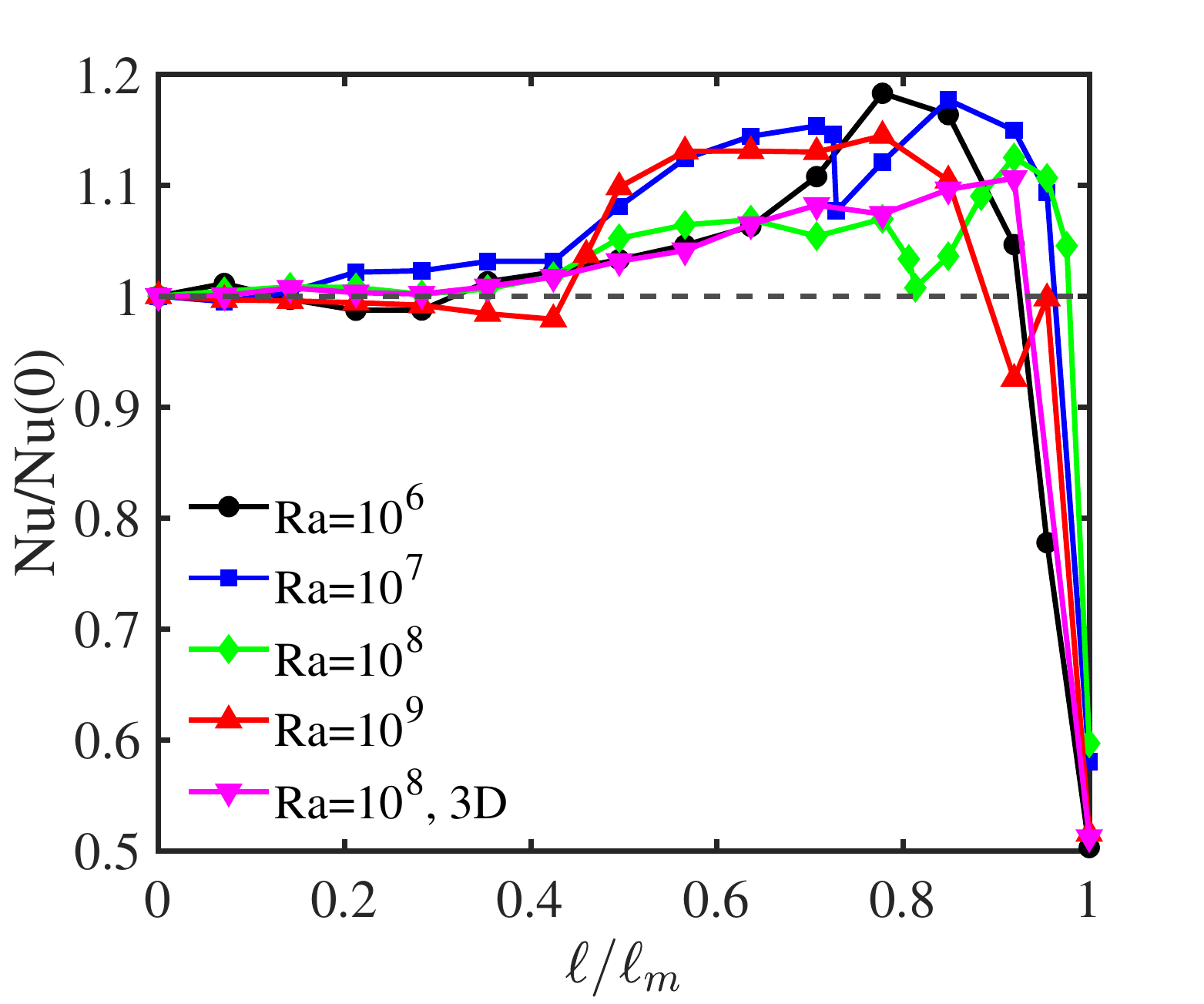}
	\put(-228,170){$(b)$}
	\\
	\includegraphics[width=0.45\linewidth]{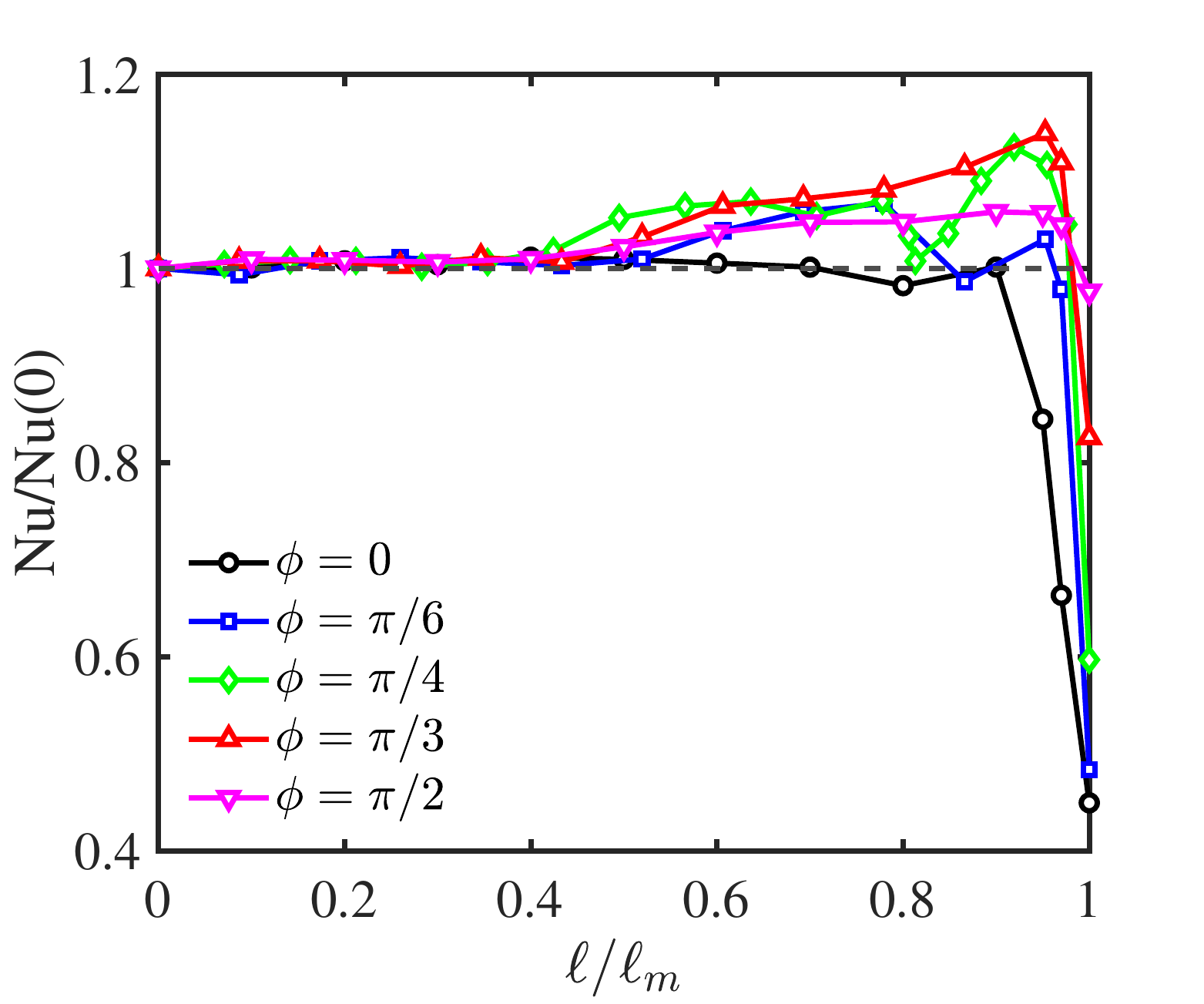}
	\put(-228,170){$(c)$}
	\includegraphics[width=0.45\linewidth]{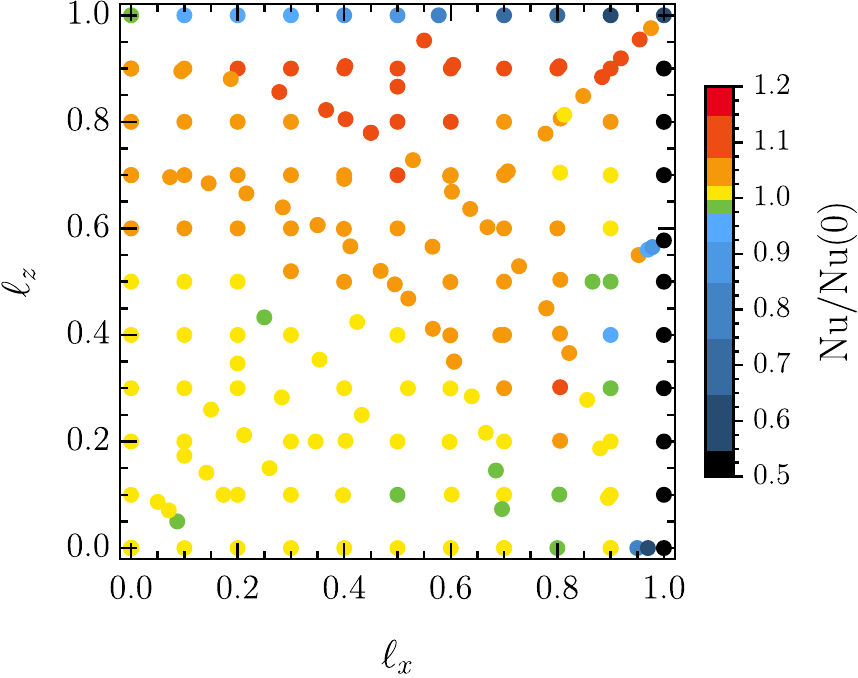}
	\put(-228,173){$(d)$}
	\\
	\includegraphics[width=0.45\linewidth]{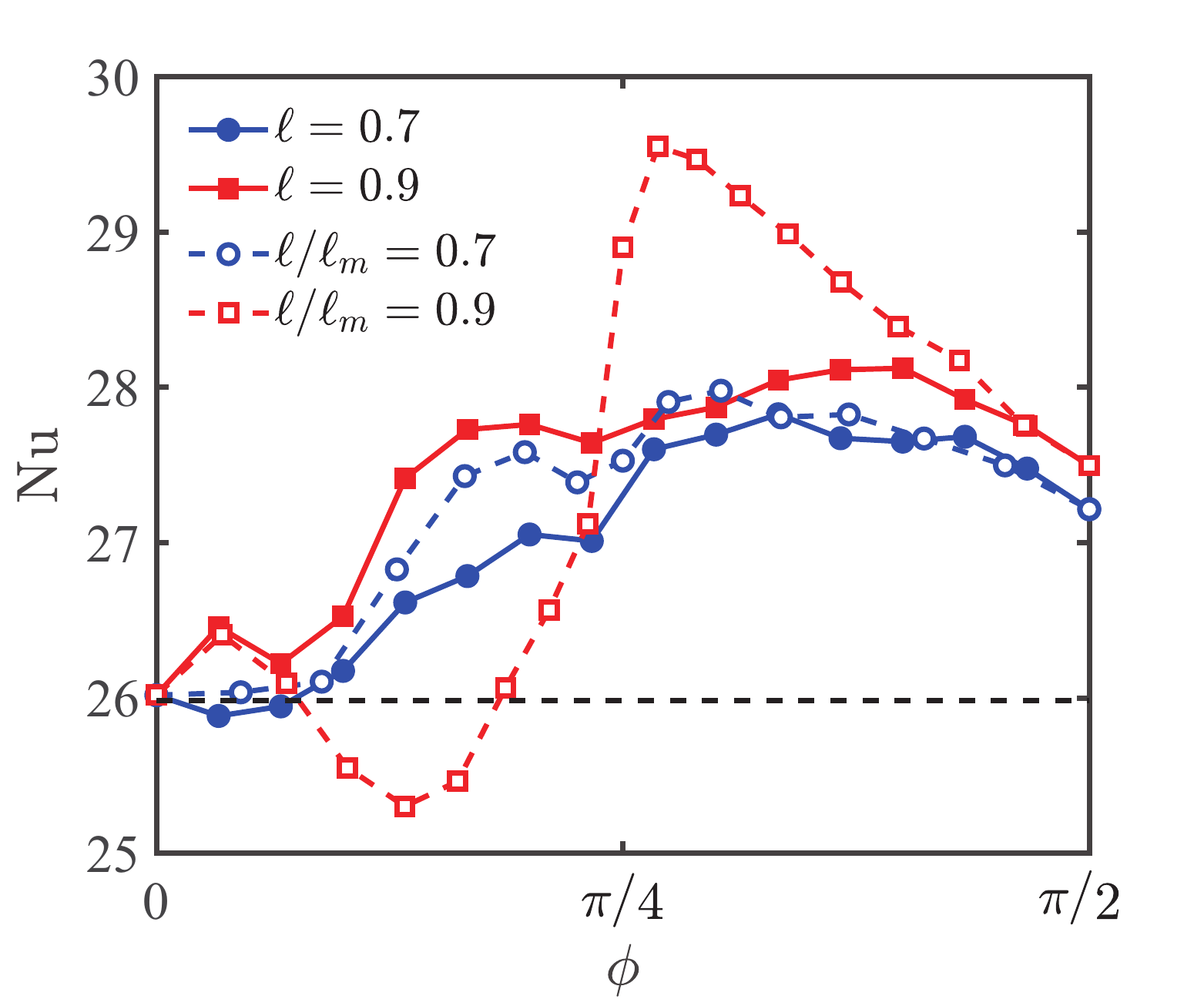}
	\put(-228,170){$(e)$}
	\vspace{-2 mm}
	\caption{\label{nu_data} $(a,b)$ Dependence of the absolute $(a)$ and normalized $(b)$ Nusselt numbers with $\ell$ for different $\Ray$ at $\phi=\pi/4$. $(c)$ Normalized Nusselt number as a function of the normalized barrier length $\ell/\ell_m$ for different $\phi$ at $\Ray=10^8$. $(d)$ Normalized Nusselt numbers in the parameter space $(\ell_x,\ell_z)$ at $\Ray=10^8$, where $\ell_{x}=\ell\cos(\phi)$ and $\ell_z=\ell\sin(\phi)$ with $(\ell,\phi)\in[0,\ell_m]\times[0,\pi/2]$. $(e)$ Dependence of $\Nus$ on angle $\phi$ for the cases $\ell=0.7$, $\ell=0.9$, $\ell/\ell_m=0.7$, and $\ell/\ell_m=0.9$, for fixed Rayleigh number $\Ray=10^8$. The value of $\Nus$ in the traditional RB convection (without barrier) is indicated by the horizontal dashed line.}
\end{figure}

We now further examine the influence of the barrier length $\ell$ on the heat transfer.
We show in Figs.~\ref{nu_data}$(a,b)$ the dependence of the absolute and normalized Nusselt numbers on $\ell$ for different $\Ray$ at a fixed barrier angle $\phi=\pi/4$.
When $\ell$ is small, the heat transfer is only mildly influenced by the barrier, as expected. As $\ell$ increases, the influences of the barrier become more significant, and heat transfer enhancement is observed for a broad range of $\ell$. The global heat transfer can be enhanced up to 18\% in the parameter space explored.
When $\ell$ is further increased to reach $\ell_m$, the top and bottom sub-regions are effectively blocked, and $\Nus$ decreases rapidly.
Similar behavior holds for different $\Ray$, showing the robustness for the barrier to enhance/reduce the heat transfer at different $\ell$.
The anomalous, non-smooth variation of $\Nus$ with $\Ray$ is associated with the change of flow structure, such as the size of the plumes in the cell corner. Further discussions on the change of corner flow with the increase of $\ell$ are presented in Appendix \ref{appB}, including visualizations of the streamlines.
Figs.~\ref{nu_data}$(a,b)$ also include the results of 3D simulations with spanwise depth 1/4. It is found that the 3D cases show similar behavior as their 2D counterparts, providing a stronger evidence that the heat transfer enhancement by adding a barrier is also relevant to physical systems.

Fig.~\ref{nu_data}$(c)$ shows the dependence of $\Nus$ with the barrier length $\ell$ for different $\phi$ at $\Ray=10^8$. Heat transfer enhancement is observed for a large range of $\phi$. However, no evident heat transfer enhancement is observed for $\phi=0$, which is attributed to the fact that when the barrier is aligned horizontally (and thus not symmetry-breaking), the interaction between the colder (hotter) fluid with the bottom (top) boundary layer is weakened by the diffusion effects during the travel over half the cell height.
In the completely blocking configuration with $\phi=0$, the system consists of two vertically stacked RB cells with height $L_b=L/2$, aspect ratio $\Gamma_b=2$ and temperature drop $\Delta_b=\Delta/2$, where the subscript $b$ indicates quantities of the completely blocking sub-systems. Thus the effective Rayleigh number of the sub-systems $\Ray_b$ is $\Ray/16$. Using the effective scaling law $\Nus\propto \Ray^{0.3}$ with the effects of aspect ratio neglected for simplicity, it is expected that $\Nus_b\approx0.44 \Nus$, which is consistent with our simulation results.
When the barrier aligns vertically with $\phi=\pi/2$ and $\ell=\ell_m$, $\Nus$ is only mildly influenced as compared to that in the traditional RB convection. In this configuration, the RB cell consists of two horizontally adjacent RB cells with aspect ratio $\Gamma_b=0.5$ and Rayleigh number $\Ray_b=\Ray$ which share a common wall and thus can also exchange heat between them.

In Fig.~\ref{nu_data}$(d)$ we display the normalized Nusselt numbers in the parameter space ($\ell_x,\ell_z$) at $Ra=10^8$, where $\ell_x=\ell\cos(\phi)$ and $\ell_z=\ell\sin(\phi)$ with $(\ell,\phi)\in[0,\ell_m]\times[0,\pi/2]$. It is found that the heat transfer is only mildly affected for small $\ell$; for barriers with $\ell\leq 0.5$ the Nusselt numbers are all within $2.5\%$ (green and yellow data points in Fig.~\ref{nu_data}$(d)$) the nominal case of no barrier. Considerable heat transfer enhancement is observed for intermediate $\ell$, and heat transfer is reduced in the completely blocking configurations for different $\phi$ (points on the top and on the right in the phase space).
Fig.~\ref{nu_data}$(d)$ shows that the influence of barrier on the heat transfer is dependent on $\phi$.
With the increase of $\phi$ above $\pi/4$, the heat transfer reduction in the completely blocking configuration is alleviated, as now the two compartments both directly connect with the top and bottom boundaries, without going through the conducting barrier. When $\phi$ is relatively small, hardly any heat transfer enhancement is observed which is consistent with the observation in Fig.~\ref{nu_data}$(c)$. 
In Fig.~\ref{nu_data}$(e)$ we show for two types of cases and for two barrier lengths the $\phi$ dependence of $\Nus$. Case one is constant barrier length $\ell$ and case two is constant $\ell/\ell_m$ which can be seen as a constant blocking fraction. For both of these cases we show two lengths (or length ratios): $\ell=0.7$, $\ell=0.9$, $\ell/\ell_m=0.7$, and $\ell/\ell_m=0.9$.
It is found that there is little heat transfer enhancement (or even a lightly reduction) for small $\phi$ and evident increase of $\Nus$ is observed at large $\phi$.
We notice that for the small barrier lengths: $\ell=0.7$, $\ell=0.9$, $\ell/\ell_m=0.7$ the trend of the flow is roughly the same for all cases. The case of large blockage fraction $\ell/\ell_m=0.9$ shows a strong $\phi$ dependence different from the other 3 cases.
As $\ell/\ell_m$ is increased from 0.7 to 0.9, the heat transfer reduction at relatively small $\phi$ $\Nus$ is attributed to the enhanced blockage, while heat transfer enhancement at large $\phi$ is attributed to the stronger perturbation of the thermal boundary layers due to the barrier.

\subsection{Flow organization}\label{sec:flow_organization}

\begin{figure}[htpb!]
	\centering
	\includegraphics[width=0.95\linewidth]{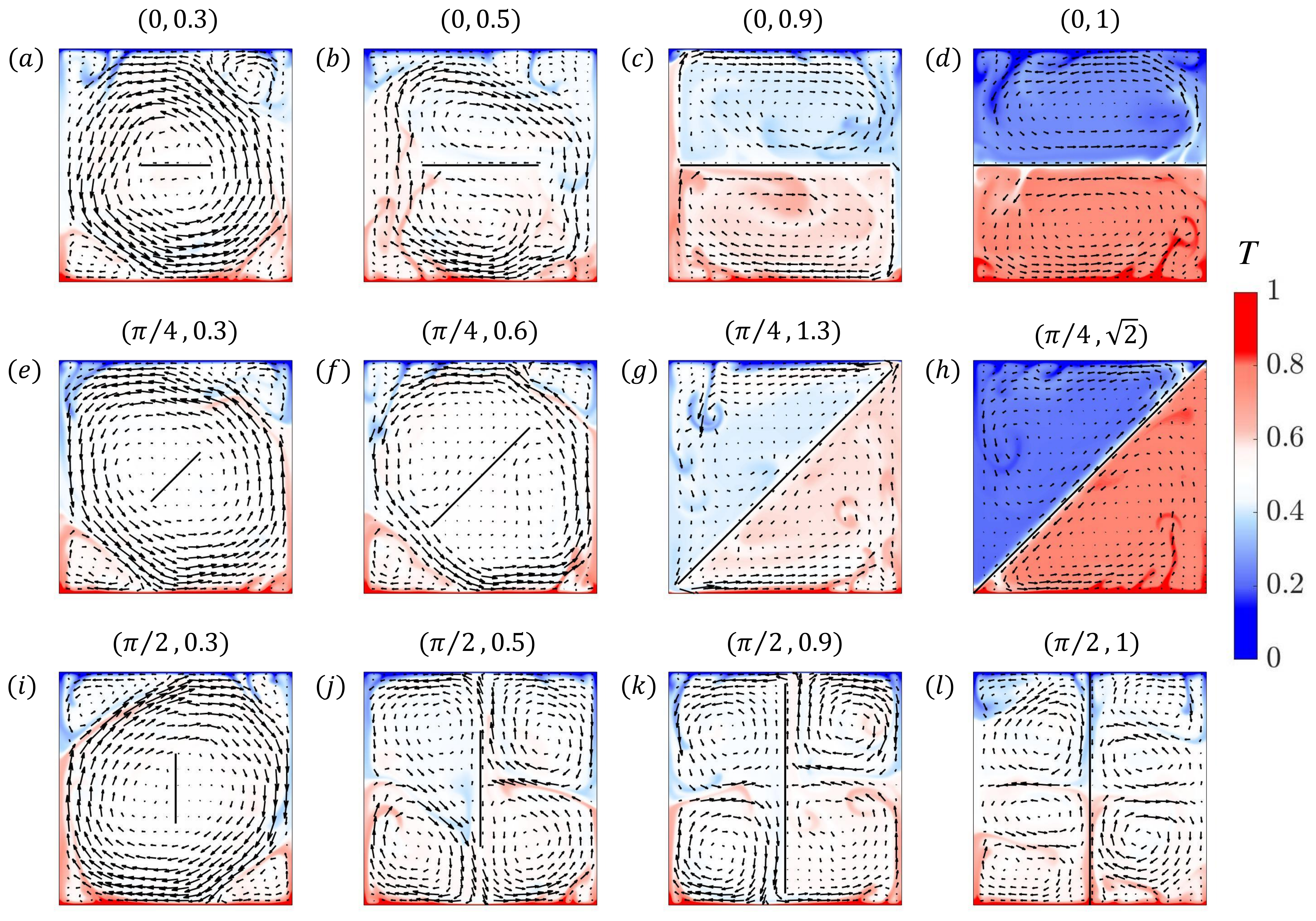}
	\vspace{-2 mm}
	\caption{\label{flow_structure}Instantaneous temperature fields superimposed with velocity vectors for various barrier lengths $\ell$ and angles $\phi$ at $\Ray=10^8$. On the top of each plot, corresponding values of $(\phi,\ell)$ are labeled. $(a-d)$ $\ell=0.3,~0.5,~0.9,~1.0$ with $\phi=0$, $(e-h)$ $\ell=0.3,~0.6,~1.3,~\sqrt{2}$ with $\phi=\pi/4$, $(i-l)$ $\ell=0.3,~0.5,~0.9,~1.0$ with $\phi=\pi/2$. In each plot the conducting barrier is denoted by a black, solid line.}
\end{figure}

We now focus on the effects of the barrier on the flow organization. Fig.~\ref{flow_structure} shows the instantaneous temperature fields superimposed with velocity vectors for various $\ell$ and $\phi$ at $\Ray=10^8$.
When a short barrier is introduced in the cell, the flow organization is close to that of the traditional RB convection. The flow consists of a well-organized LSC and two corner rolls, see e.g.~Figs.~\ref{flow_structure}$(a,e,i)$. The temperature in the bulk is well mixed, and the temperature drop is dominated by the thermal boundary layers near the horizontal plates.
The barrier is located in the center of the LSC, where the flow velocity is low and temperature is uniform.
It is expected that a short barrier has only a mild influence on the heat transfer, as shown in \S\ref{sec:heat_transfer}.

Although the flow structure is qualitatively unaffected by a short barrier, we can see that a short barrier can obviously distort the LSC and influence the competition between the LSC and the corner rolls. Besides, an inclined barrier will break the parity symmetry of the system with respect to the horizontal mid-plane. Thus, flow reversal process is expected to be influenced by the barrier, as shown in \S\ref{sec:reversal}.

\begin{figure}[htpb!]
	\centering
	\includegraphics[width=0.45\linewidth]{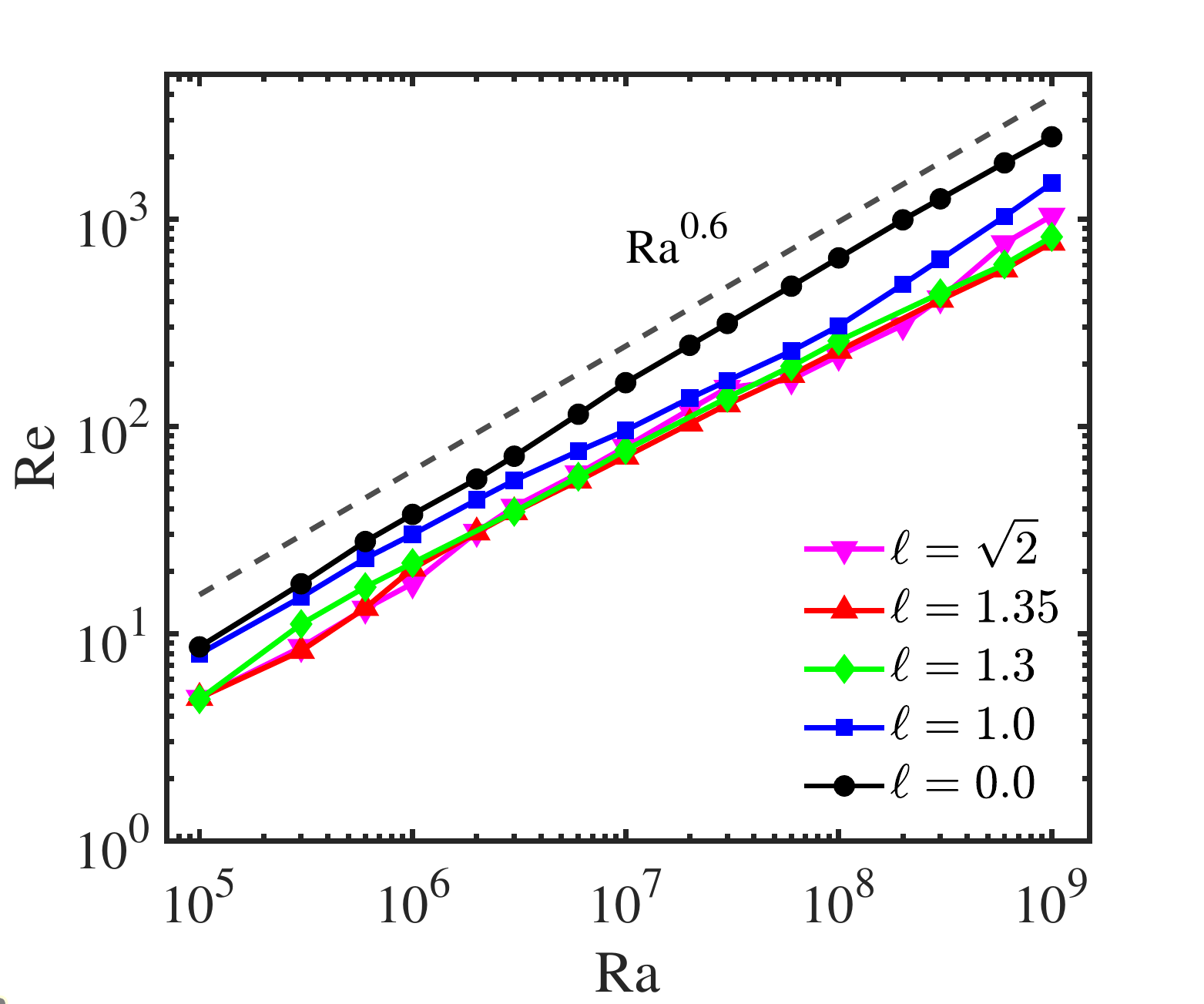}
	\put(-228,170){$(a)$}
	\includegraphics[width=0.45\linewidth]{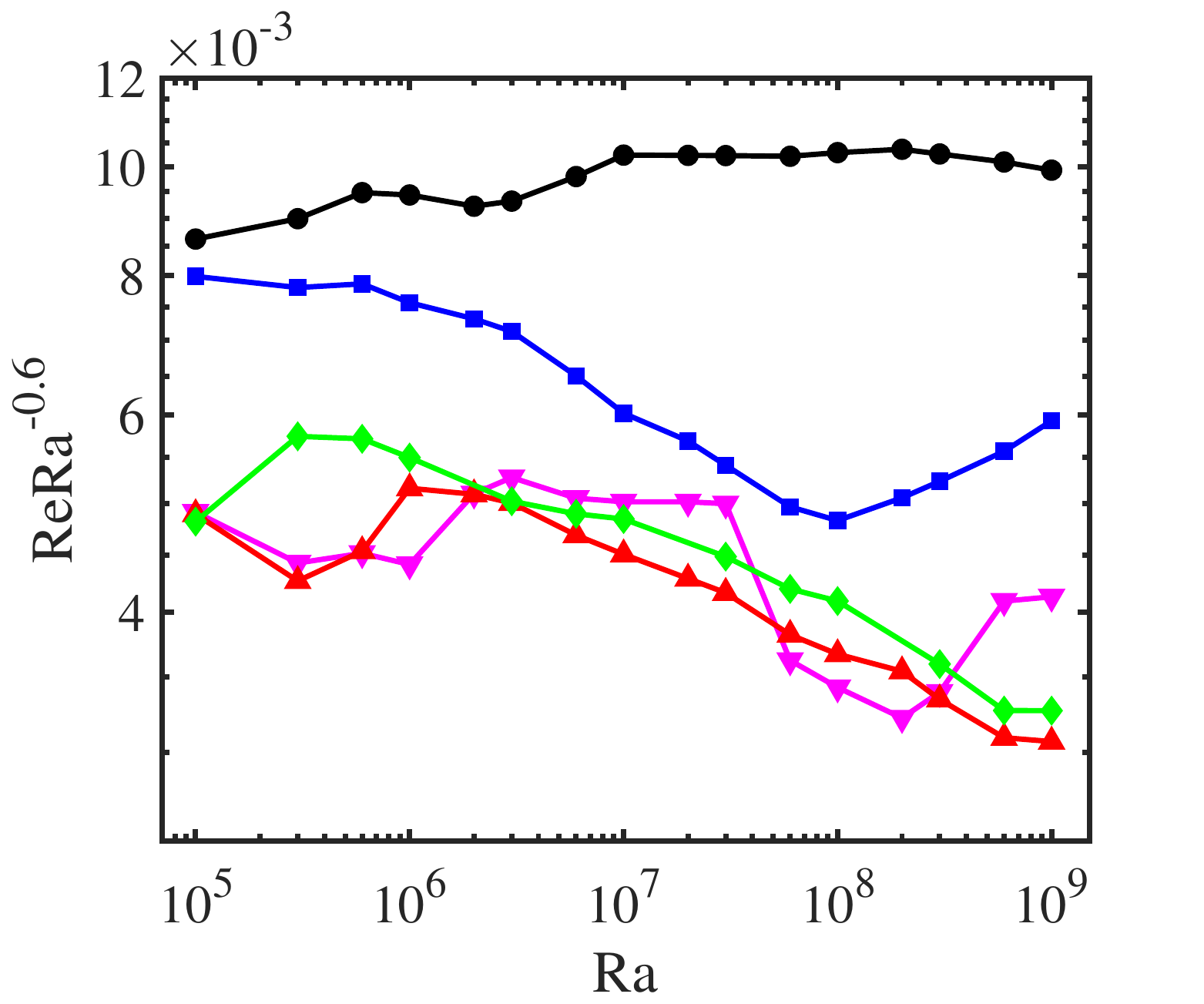}
	\put(-228,170){$(b)$}
	\vspace{-2 mm}
	\caption{\label{re_ra_scaling} $(a)$ $\Rey$ as a function of $\Ray$ for different $\ell$ for a fixed barrier angle $\phi=\pi/4$. For reference, a $\Rey \propto \Ray^{0.6}$ scaling is included as a dashed grey line. $(b)$ The compensated plot of $(a)$ revealing the details of the overlapping lines in $(a)$.}
\end{figure}

When the barrier becomes longer, the flow organization is significantly modified. It is appropriate to regard the convection cell to be the combination of two sub-regions with mass and heat exchange between each other. Complex vortex structures develop in each sub-region, as depicted in Figs.~\ref{flow_structure}$(b,f,j)$.
When the barrier becomes even larger, the gaps connecting the two sub-regions become smaller. Due to the suppressed mass transfer between the two sub-regions, the fluid mixing in the cell becomes less efficient. For $\phi<\pi/2$, the fluid in the top (bottom) sub-region is colder (hotter) due to the imbalance between the cold and hot boundaries in each sub-region, suggesting that a temperature boundary layer gradually develops on the barrier surface, see Figs.~\ref{flow_structure}$(c,g)$.
Note that strong jets develop across the small gaps, which are attributed to the convergence of fluid flow and the pressure imbalance near the gaps, which originates from the distinct temperature values in the two sub-regions.

The interaction of the jets with the boundary layers can promote plume generation, which is beneficial for the heat transfer \cite{blass2020}.
For $\phi=0$, the jets travel half the cell height before interacting with the boundary layers, in contrast to the cases for larger $\phi$.
The phenomena of jet formation and the resulting heat transfer enhancement are similar to those found in the partitioned RB convection \cite{bao2015}. When vertical partitions are inserted in the convection system, coherent convective flow is established spontaneously, with the hotter and cooler fluid moving upwards and downwards, respectively. Due to the horizontal pressure drops across the thin gaps between the partitions and horizontal plates, strong horizontal jets form and sweep the thermal boundary layers, realizing efficient heat transfer.

When $\ell=\ell_m$, the two sub-regions are completely separated without mass transfer between each other. Heat is transferred between the two sub-regions via thermal conduction through the barrier. For $\phi<\pi/2$, the averaged temperature values in the two sub-regions are different, particularly for small $\phi$. When $\phi$ is relatively small, the temperature difference on the two sides of the barrier is significant and a thermal boundary layer appears on the barrier, which can also generate thermal plumes, as shown in Figs.~\ref{flow_structure}$(d,h)$.
See also movie S4 in the Supplemental Material for an illustration of the flow dynamics in the completely blocking configuration with $\phi=\pi/4$.
Note that the barrier remains conductive for all cases with the same thermal properties as the fluid, such that, even for the cases of complete blockage $\ell=\ell_m$ and $\phi\le\pi/4$, there is a net thermal transport from the bottom plate to the top plate.
For $\phi>\pi/4$, each of the two sub-regions contains part of the top and bottom plates, and heat can be directly transferred between the two plates through thermal convection besides of conduction. When $\phi=\pi/2$, the two sub-regions are similar to the traditional RB convection in a cavity with aspect ratio $\Gamma_b=1/2$.
For $\phi\le\pi/4$, heat is transferred across the barrier through thermal conduction.

When the barrier is put in the cell, it is expected that the flow strength will be reduced due to the additional drag exerted by the barrier on the flow. We now quantitatively examine how the flow strength is affected by the barrier.
The flow strength is measured by the volume-averaged Reynolds number $\Rey$, which is defined as
\begin{equation}
\Rey=L\sqrt{\frac{\Ray}{\Pra}} \sqrt{\langle\vec{u}\cdot\vec{u}\rangle_{V,t}}
\end{equation}
where $\left\langle\hdots\right\rangle_{V,t}$ denotes averaging the bracketed value over the entire volume and time. Fig.~\ref{re_ra_scaling}$(a)$ shows the dependence of $\Rey$ on $\Ray$ for different $\ell$ for the case of $\phi=\pi/4$.
In the traditional RB convection with $\ell=0$, $\Rey$ follows an effective scaling law, $\Rey \propto \Ray^{0.6}$.
When the barrier is introduced, $\Rey$ is uniformly decreased for all $\Ray$ under consideration, as expected. 
The scaling behavior of $\Rey$ with $\Ray$ is only mildly modified in the presence of barrier despite of the remarkable changes of flow organization and strength. See also the compensated figure showing $\Rey \Ray^{-0.6}$ in Fig.~\ref{re_ra_scaling}$(b)$, where the details of the overlapping lines can be seen.
It is interesting to compare Fig.~\ref{re_ra_scaling} with the behavior of $\Nus$ in Fig.~\ref{nu_ra_scaling}. While in the completely blocking configuration with $\ell=\ell_m$, both $\Nus$ and $\Rey$ are significantly reduced compared to those in the traditional RB convection, $\Nus$ is very robust compared to $\Rey$ when the barrier has a length $\ell<\ell_m$, particularly for high $\Ray$. 
The slight decrease of $\ell$ from $\ell_m$ results in a mild change of flow strength and a huge heat transfer enhancement, demonstrating the effect of jets to enhance the heat transfer.

\begin{figure}[htpb!]
	\centering
	\includegraphics[width=0.45\linewidth]{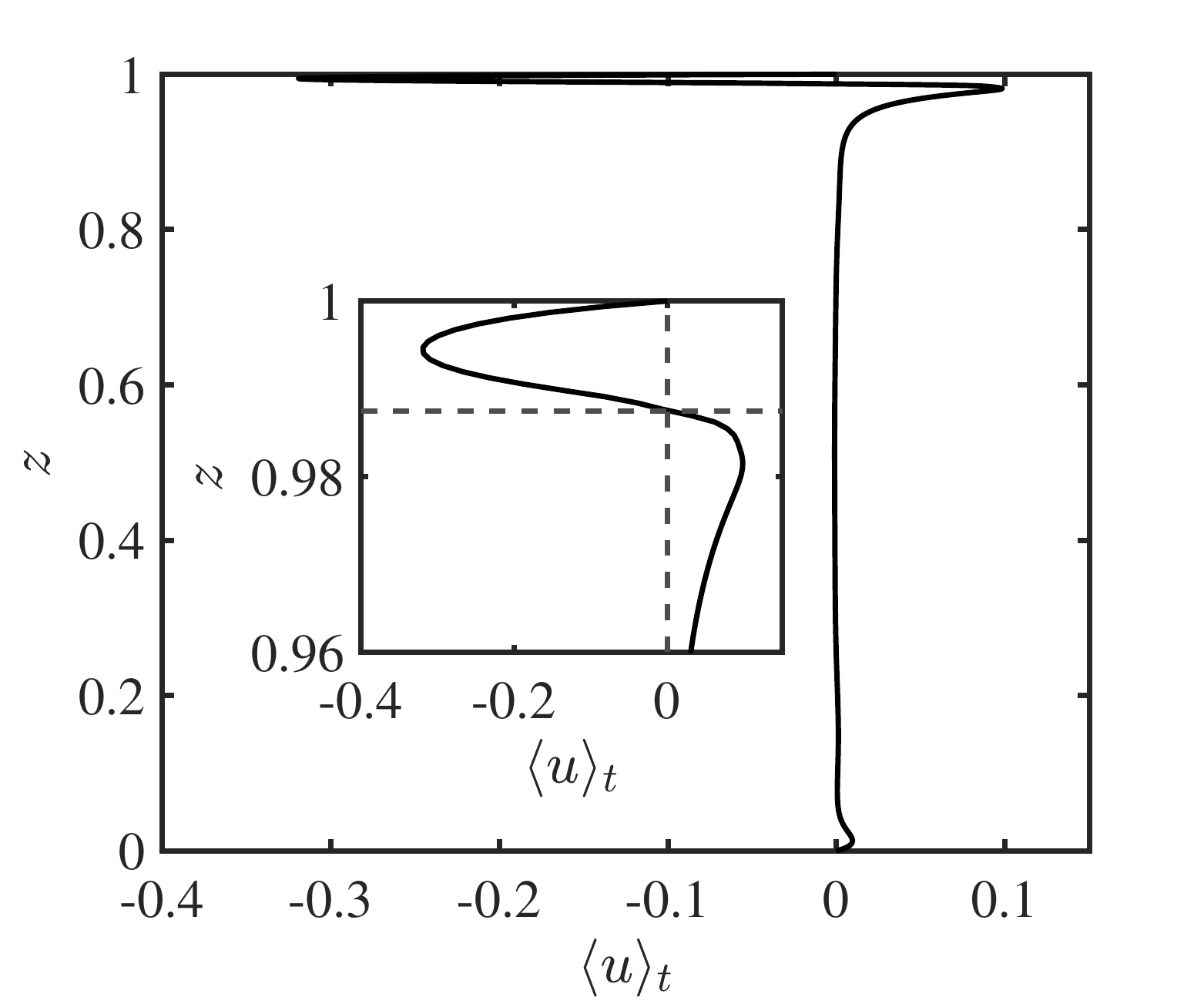}
	\put(-228,170){$(a)$}
	\includegraphics[width=0.48\linewidth]{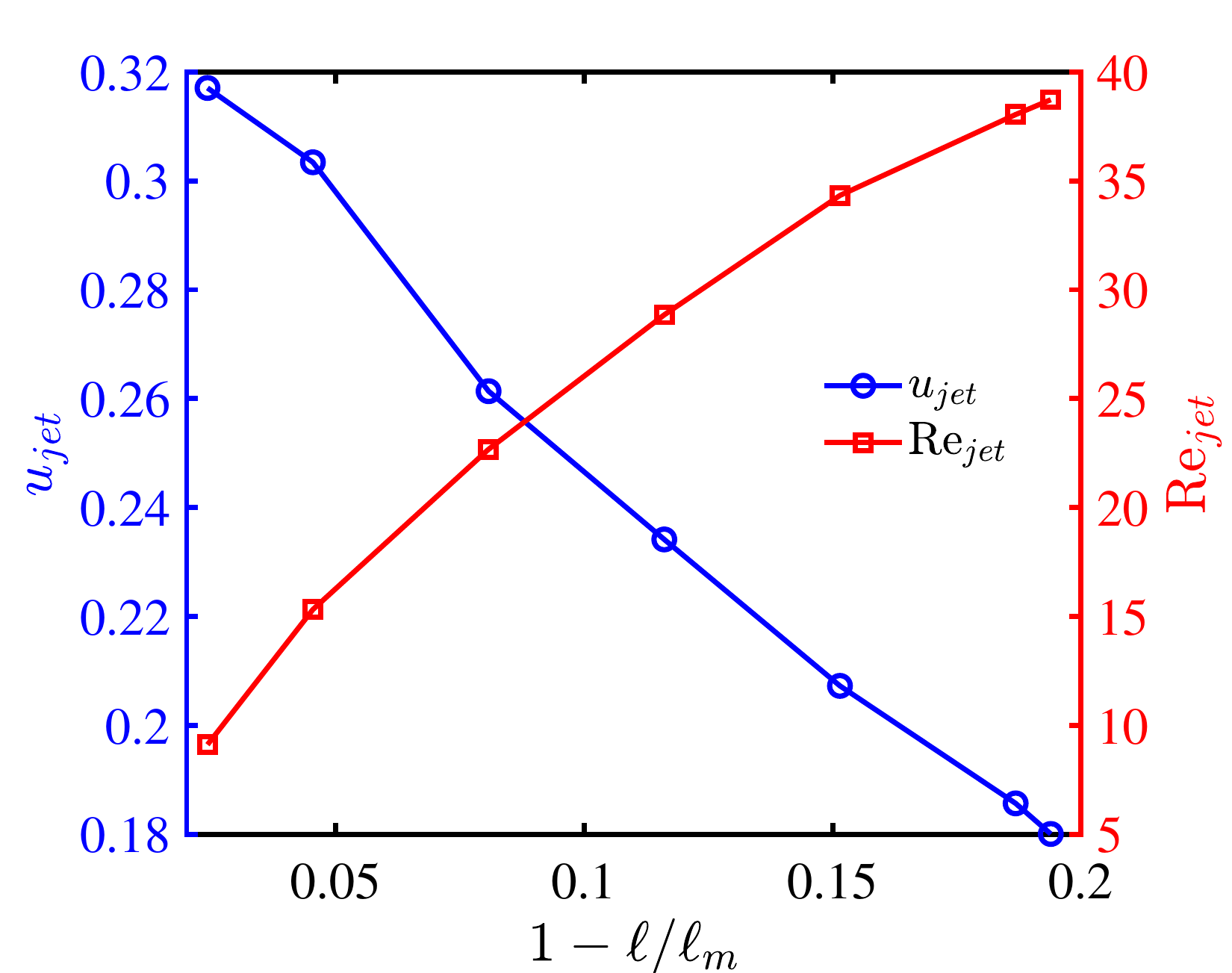}
	\put(-242,170){$(b)$}
	\vspace{-2 mm}
	\caption{\label{velo_slice_barr_end}$(a)$ Profile of the time-averaged horizontal velocity $\langle u\rangle_t(z)$ along the vertical slice crossing the barrier end at $\Ray=10^8$, with a barrier angle $\phi=\pi/4$ and a barrier length $\ell=1.38$. Inset: Zoom of the same data. The vertical position of the barrier tip is indicated by the horizontal dashed line. $(b)$ The maximum velocity of jet $u_{jet}$ and the jet Reynolds number $\Rey_{jet}$ as functions of $1-\ell/\ell_{m}$ for $Ra=10^8$ and $\phi=\pi/4$.}
\end{figure}

To further investigate the jet structure, we show in Fig.~\ref{velo_slice_barr_end}$(a)$ the profile of the time-averaged horizontal velocity $\langle u\rangle_t(z)$ along the vertical slice crossing the barrier end at $\Ray=10^8$, with a barrier angle $\phi=\pi/4$ and a barrier length $\ell=1.38$, where $\langle\hdots\rangle_t$ indicates averaging over time. The existence of jet structure is clearly demonstrated and the jet is found to approximately take a parabolic profile. Fig.~\ref{velo_slice_barr_end}$(b)$ shows the maximum value $u_{jet}$ of the jet velocity $|\langle u\rangle_t|$ as a function of $1-\ell/\ell_m$. For the parameters explored, $u_{jet}$ increases for increasing $\ell$ (except for the case $\ell=\ell_m$), showing the formation process of the jet. When $\ell$ is large enough with $Ra$ fixed, it is expected that $u_{jet}$ will decrease with $\ell$ and vanishes at $\ell=\ell_m$. We quantify the width of the jet using the distance $\delta_{jet}$ of the extreme point of velocity from the horizontal plate, which is found to decrease with decreasing gap width. The capability of mass transfer by the jet is measured using the jet Reynolds number $\Rey_{jet}=u_{jet}\delta_{jet}\sqrt{Ra/Pr}$, which is shown in Fig.~\ref{velo_slice_barr_end}$(b)$ as a function of $1-\ell/\ell_m$. It is found that $\Rey_{jet}$ decreases with increasing $\ell$, indicating that the mass transfer between the two sub-regions is suppressed as the gap is narrowed, even though a strong jet develops.

\begin{figure}[htpb!]
	\centering
	\hspace{-6 mm}
	\includegraphics[width=0.45\linewidth]{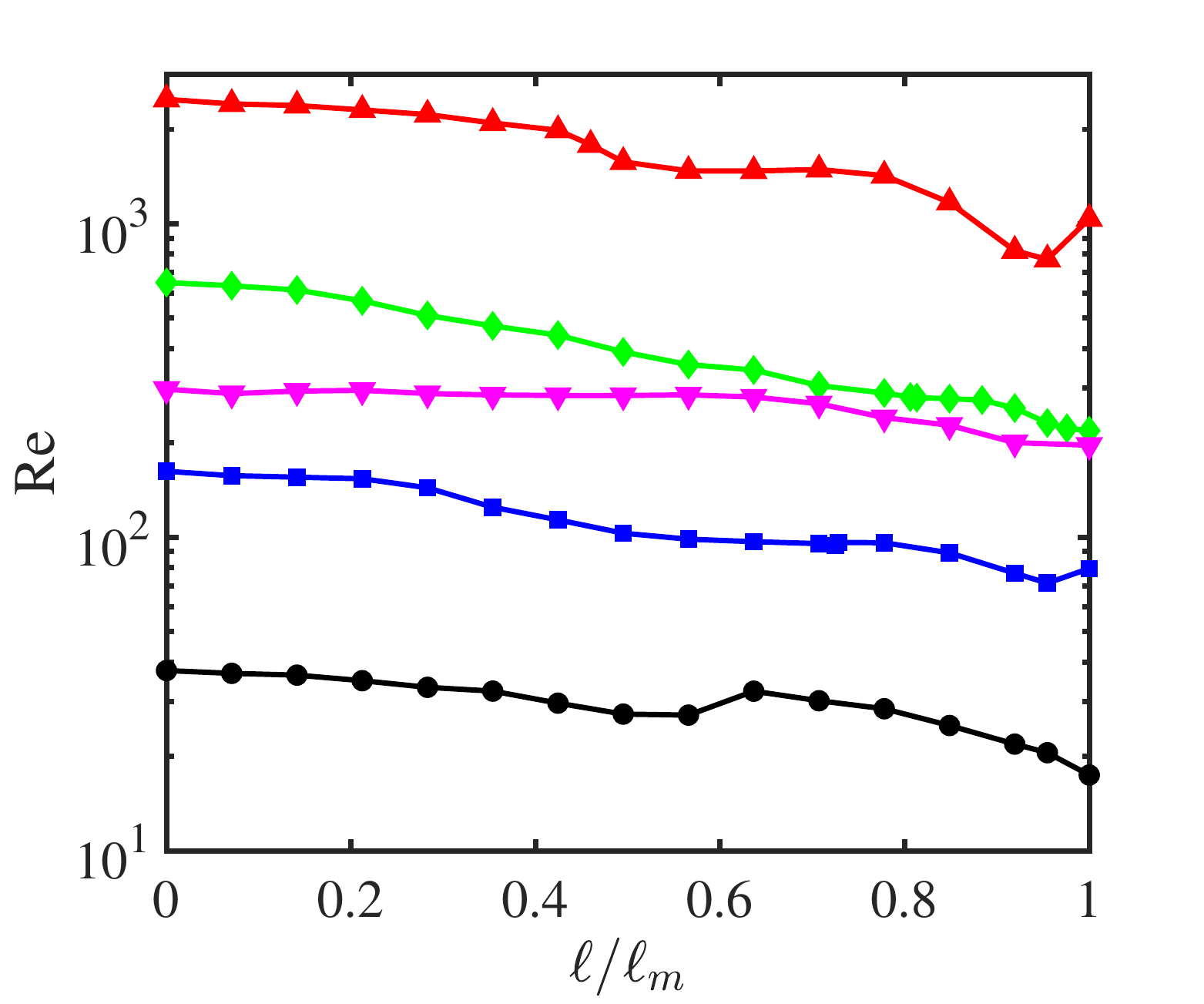}
	\put(-224,170){$(a)$}
	\includegraphics[width=0.45\linewidth]{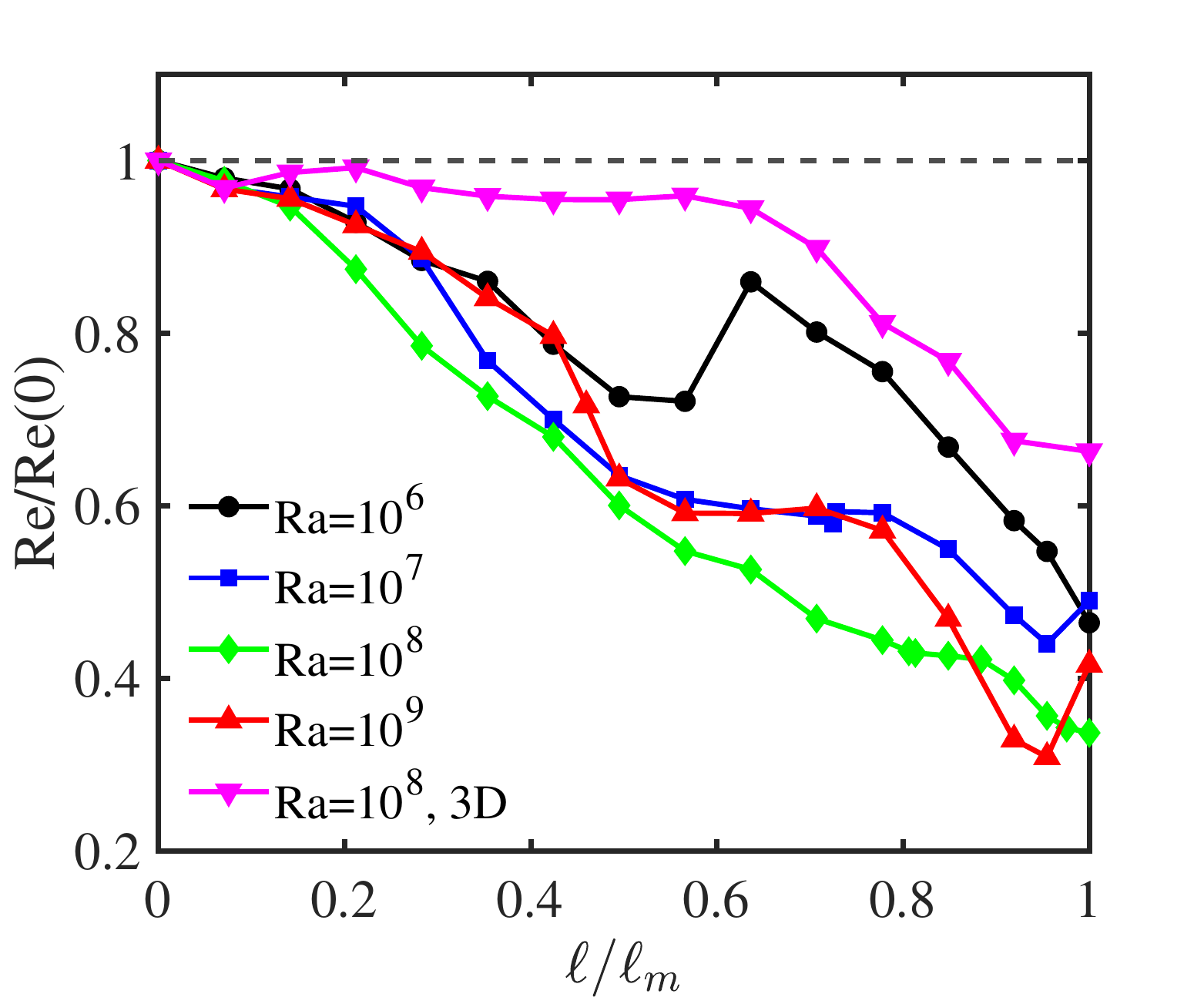}
	\put(-224,170){$(b)$}
	\\
	\hspace{-6 mm}
	\includegraphics[width=0.45\linewidth]{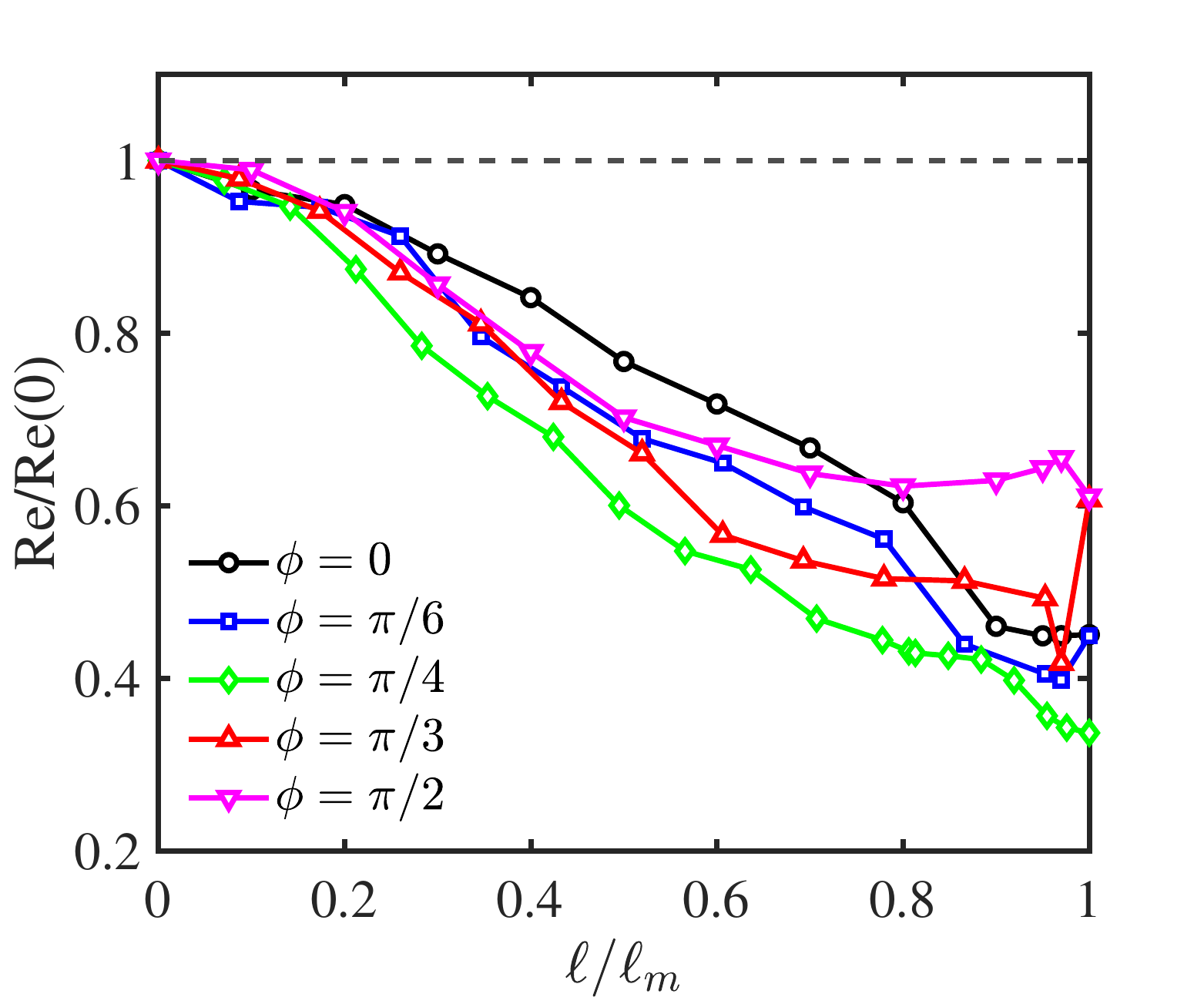}
	\put(-224,170){$(c)$}
	\includegraphics[width=0.45\linewidth]{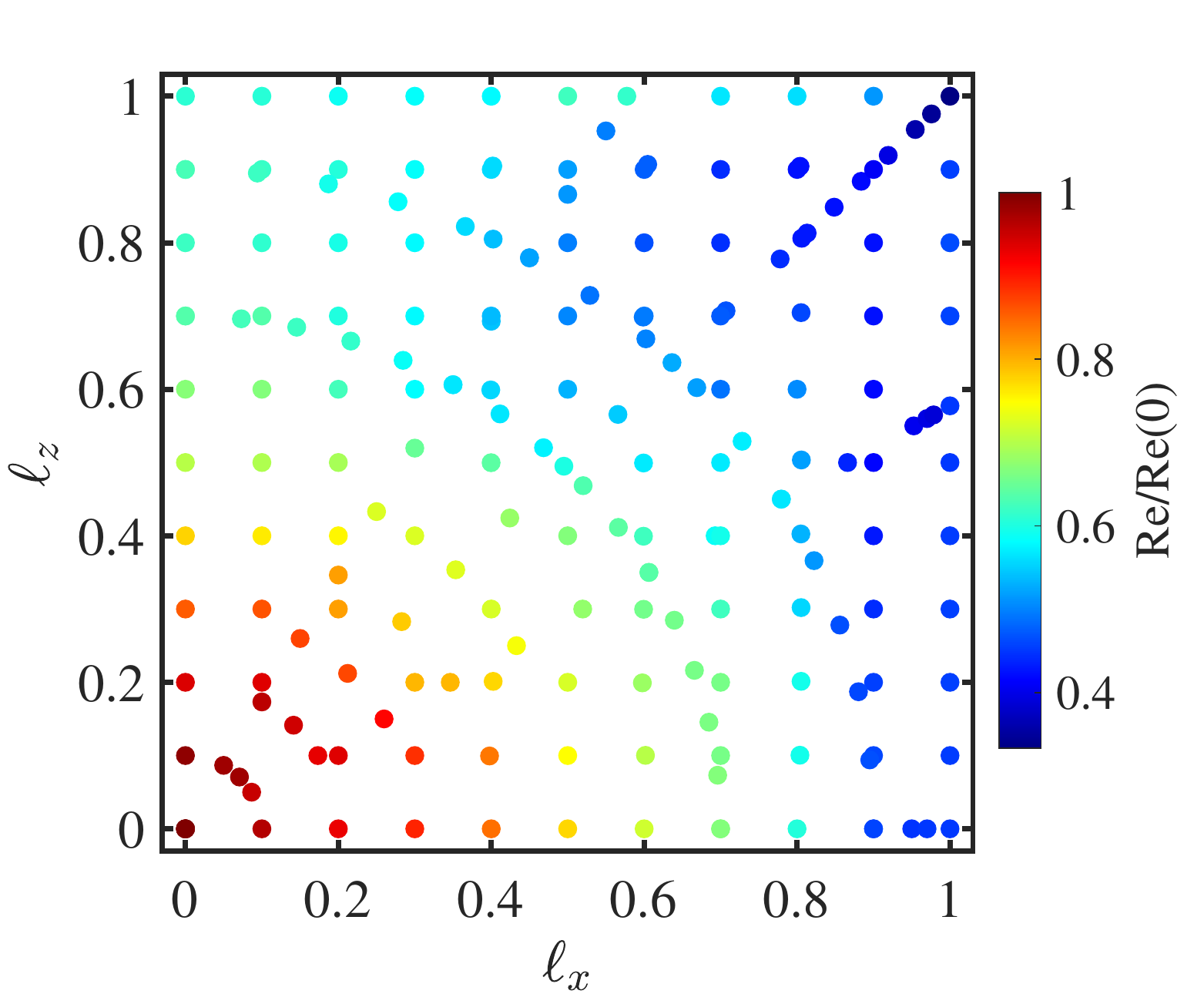}
	\put(-228,170){$(d)$}
	\vspace{-2 mm}
	\caption{\label{re_data} $(a,b)$ Dependence of the absolute $(a)$ and normalized $(b)$ Reynolds numbers with $\ell$ for different $\Ray$ at $\phi=\pi/4$. $(c)$ Normalized Reynolds number as a function of the normalized barrier length $\ell/\ell_m$ for different $\phi$ at $\Ray=10^8$. $(d)$ Normalized Reynolds numbers in the parameter space $(\ell_x,\ell_z)$ at $\Ray=10^8$, where $\ell_{x}=\ell\cos(\phi)$ and $\ell_z=\ell\sin(\phi)$ with $(\ell,\phi)\in[0,\ell_m]\times[0,\pi/2]$.}
\end{figure}

We now focus on how the barrier influences the global Reynolds number. We display in Figs.~\ref{re_data}$(a,b)$ the absolute and normalized Reynolds numbers as functions of $\ell$ for different $\Ray$ at a fixed barrier angle $\phi=\pi/4$, showing that the flow strength is reduced in the presence of a barrier.
With larger $\ell$ the flow strength generally becomes smaller, which is attributed to the fact that, for larger $\ell$ the drag force becomes stronger and the buoyancy force acting on the plumes is reduced when the temperature mixing is less efficient.
Non-monotonic variation of $\Rey$ with $\ell$ is also observed, which is associated with the change of flow organization, such as the modification of corner flow demonstrated in Appendix \ref{appB}.
It is interesting that, in the presence of a barrier, the heat transfer can be enhanced for a broad range of $\ell$, even though the flow strength is significantly reduced. The heat transfer enhancement can therefore be attributed to the well-organized, direction-oriented motion (funneling) of the plumes and the direct interaction of the jets with the boundary layers.
In Figs.~\ref{re_data}$(a,b)$ the results of a set of 3D simulations with spanwise depth 1/4 are also included, showing a similar behavior. In the 3D case, thanks to the strong friction of the cell walls, a relatively short barrier has only a mild influence on the flow strength. Thus, $\Rey$ is insensitive to the increase of $\ell$ when $\ell$ is relatively small; while when $\ell$ becomes large, the influence of the barrier becomes significant, and $\Rey$ is found to drop rapidly with the further increase of $\ell$.

Fig.~\ref{re_data}$(c)$ shows the dependence of $\ell$ on $\Rey$ for varying $\phi$ at $\Ray=10^8$, showing that the flow strength is reduced for all $\phi$ and for all $\ell$. Fig.~\ref{re_data}$(d)$ shows the normalized Reynolds numbers in the parameter space $(\ell_x,\ell_z)$, again showing the reduced flow strength in the presence of a barrier. In the completely blocking case, the reduction of the flow strength is found to be less significant at larger $\phi$. It is attributed to the larger buoyancy force acting on the plumes due to the larger effective Rayleigh numbers $\Ray_b$.

\begin{figure}
	\centering
	\includegraphics[width=0.45\linewidth]{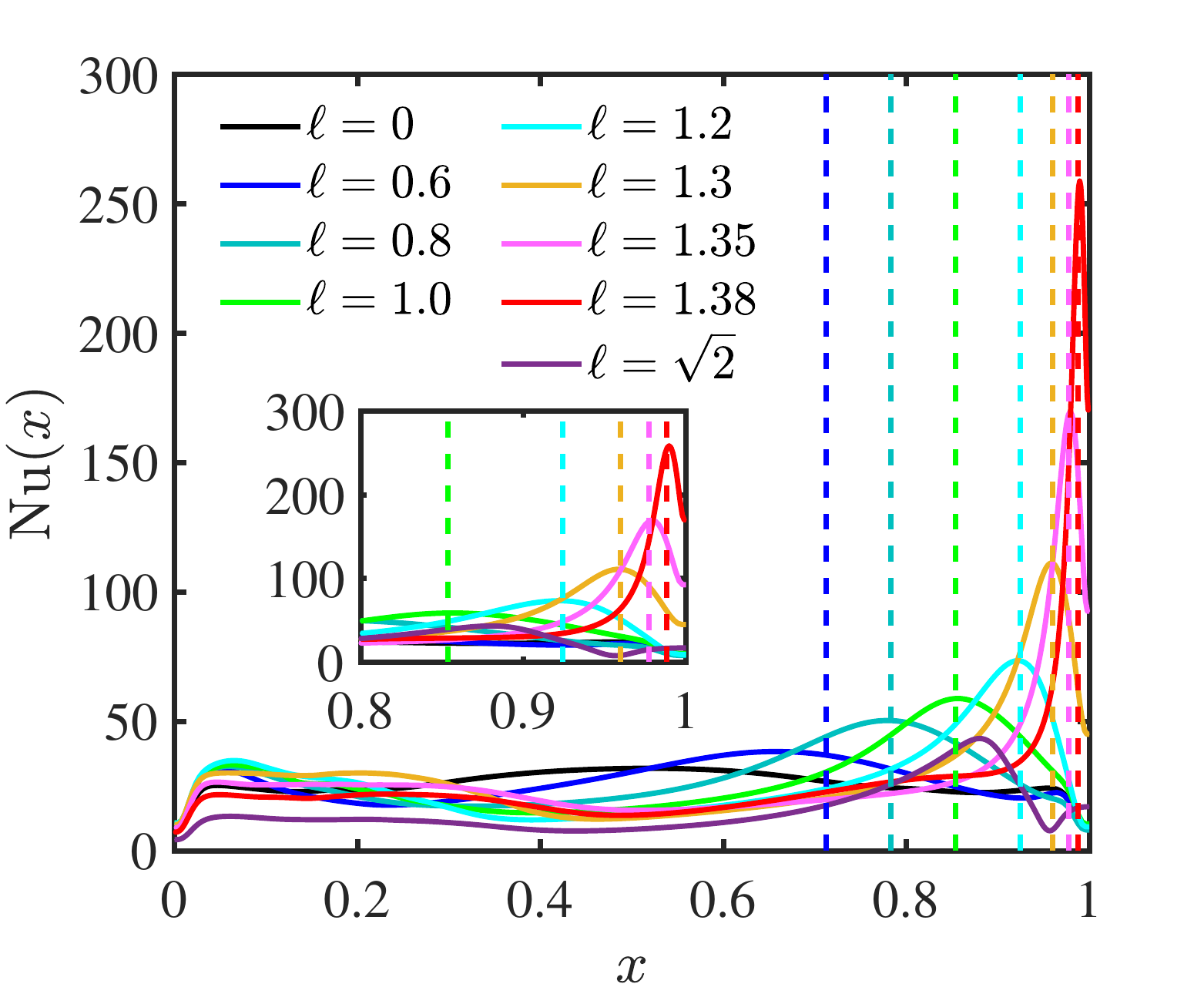}
	\put(-229,170){$(a)$}
	\includegraphics[width=0.45\linewidth]{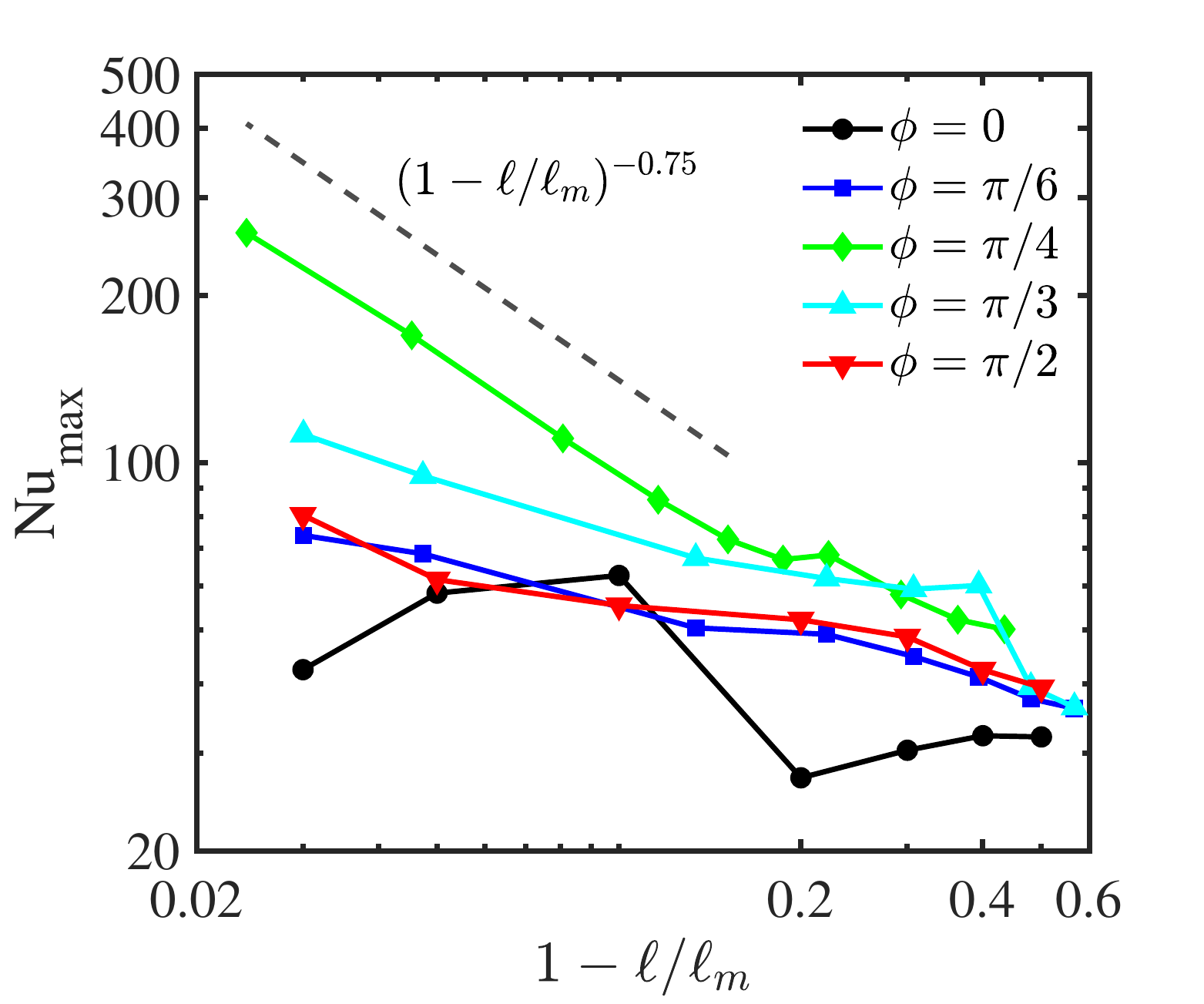}
	\put(-229,170){$(b)$}
	\vspace{-2 mm}
	\caption{\label{local_flux}$(a)$ Profiles of the time-averaged local heat flux $\Nus(x)=-\partial_z\langle T\rangle_t$ at the top plate for different barrier lengths $\ell$ at $\Ray=10^8$ and $\phi=\pi/4$. The vertical dashed lines indicate the $x$-coordinate of the right barrier end, showing that the extreme point of $\Nus(x)$ is close to the barrier end. Inset: Zoom of the same data. $(b)$ Dependence of the maximum local Nusselt number $\Nus_{\text{max}} = \max[\Nus(x)]$ at the top plate on $1-\ell/\ell_m$ for different $\phi$ for the case of $\Ray=10^8$. For reference, a $\Nus_{\text{max}}\propto(1-\ell/\ell_m)^{-0.75}$ scaling is included as a dashed grey line.
	}
\end{figure}

Considering that the present flow is in the `classical' regime of turbulent RB convection and the heat transfer is limited by the thermal boundary layers, where heat is mainly transferred via thermal conduction, it is interesting to study how the barrier influences the properties of the thermal boundary layers.
To show that, we plot in Fig.~\ref{local_flux}$(a)$ the time-averaged local heat flux $\Nus(x)=-\partial_z\langle T\rangle_t$ at the top plate for different $\ell$ at $\Ray=10^8$ and $\phi=\pi/4$. The horizontal coordinates of the barrier ends with $0<\ell<\ell_m$ are indicated by dashed lines.
It is found that as $\ell$ increases, $\Nus(x)$ increases rapidly above the barrier end, which is attributed to the direct impact of hot fluid to the top boundary layer.
Similar phenomenon appears at the bottom plate due to the rotational symmetry of the system.
For the parameters considered, the maximum $\Nus_{\text{max}}$ of $\Nus(x)$ is enhanced up to around 800\%. The results imply that the thermal boundary layers at the horizontal plates are highly nonuniform in the presence of a long barrier at $\phi=\pi/4$.
Fig.~\ref{local_flux}$(b)$ shows the dependence of $\Nus_{\text{max}}$ with the barrier length for different $\phi$. It is found that $\Nus_{\text{max}}$ increases as $1-\ell/\ell_m$ approaches 0 ($\ell$ approaches $\ell_m$), except for the case of $\phi=0$, where the interaction between the jet and thermal boundary layer is less significant than for the other cases. The growth rate of $\Nus_{\text{max}}$ with decreasing $1-\ell/\ell_m$ is dependent on $\phi$ and is always slower than $(1-\ell/\ell_m)^{-1}$. Thus, for $\phi\ge\pi/4$, the decreasing rate of the local thermal boundary layer thickness is slower than the gap width as $\ell$ increases. For large enough $\ell$, the thermal boundary layer thickness will become larger than the gap width, and the system gradually approaches the completely blocking configuration with increasing $\ell$.

\begin{figure}
	\centering
	\includegraphics[width=0.45\linewidth]{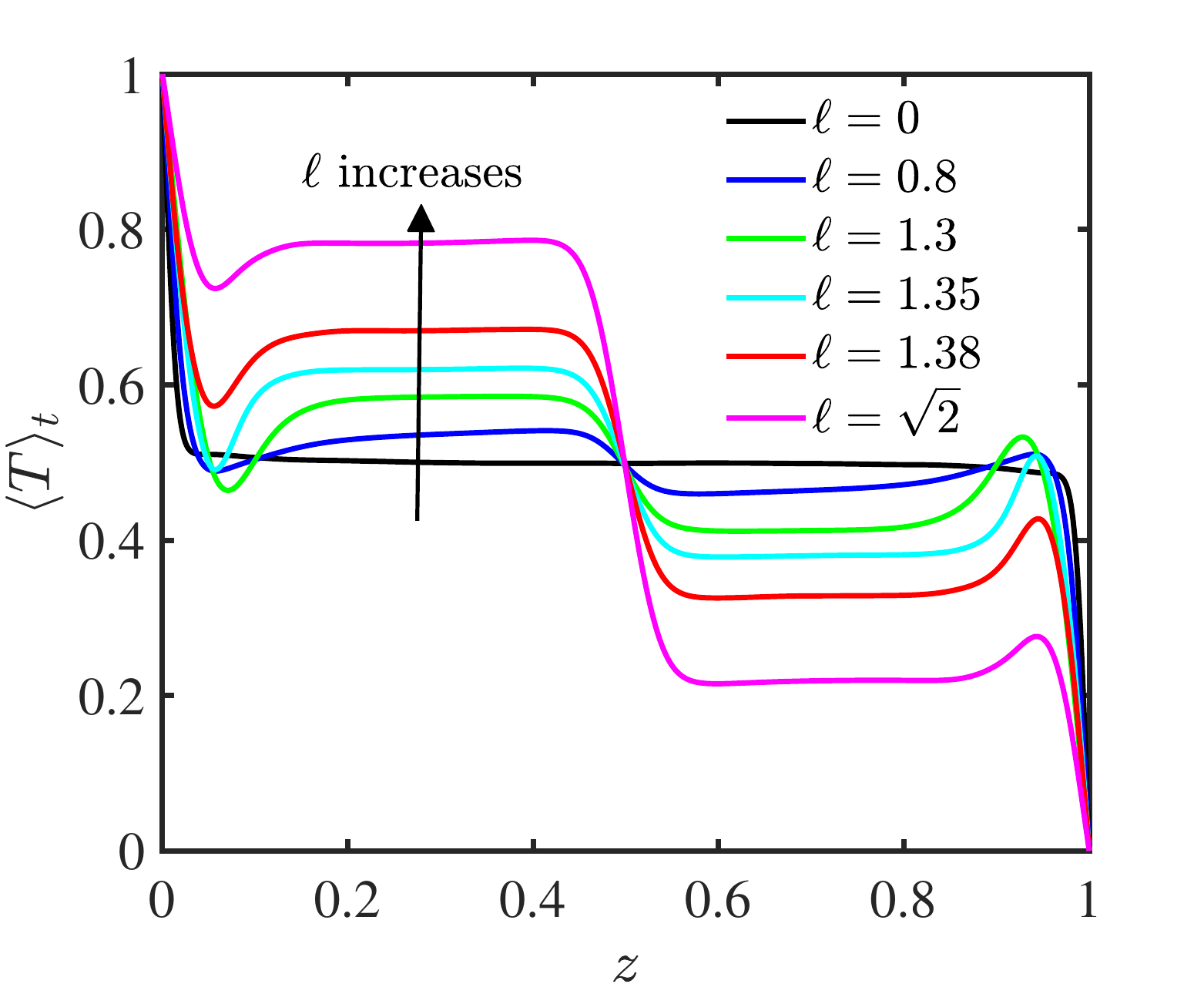}
	\put(-224,170){$(a)$}
	\includegraphics[width=0.45\linewidth]{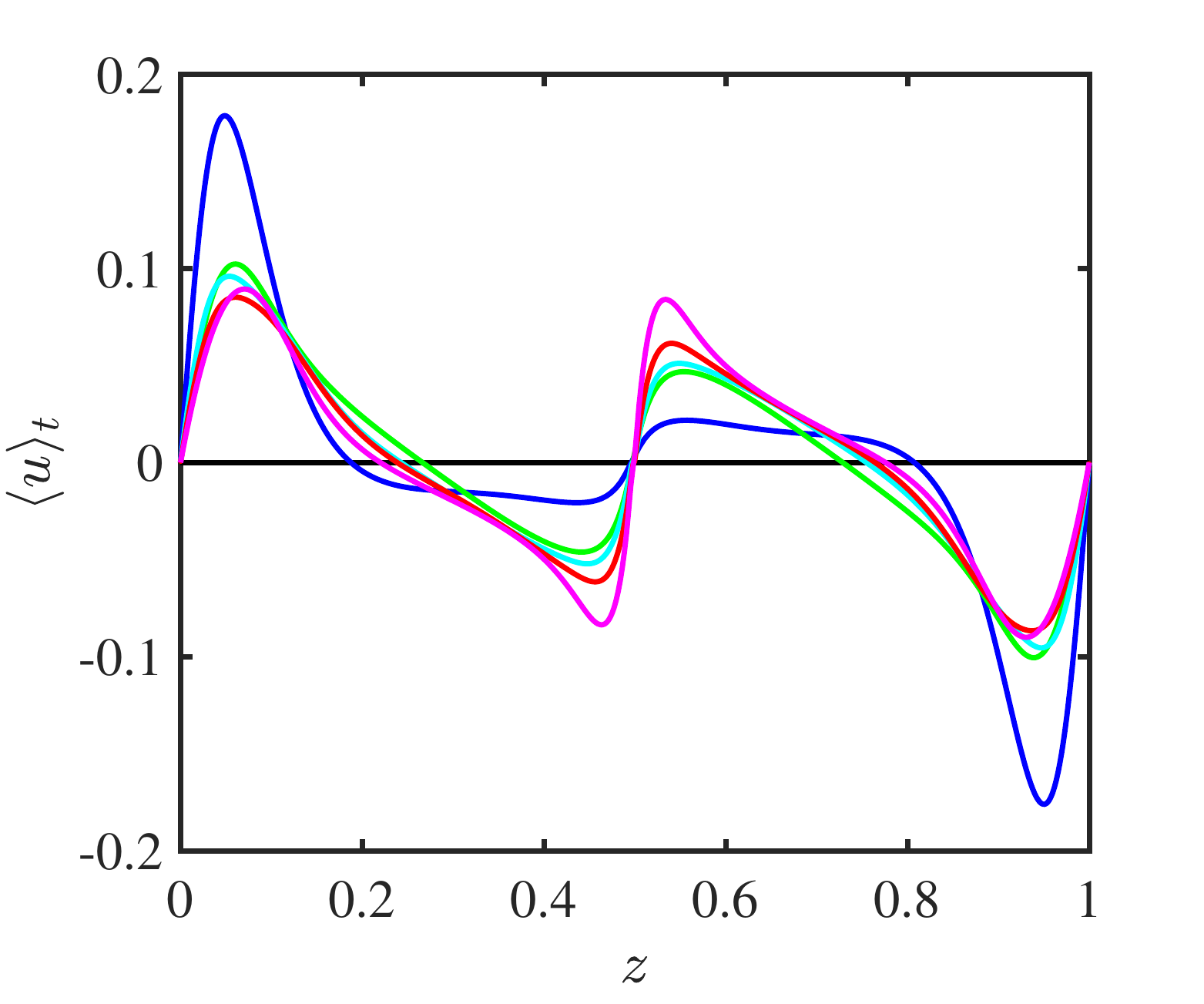}
	\put(-224,170){$(b)$}
	\vspace{-2 mm}
	\caption{\label{mean_flow_slice} Profiles of the time-averaged temperature $\langle T\rangle_t$ $(a)$ and horizontal velocity $\langle u\rangle_t$ $(b)$ at the horizontal mid-plane for different barrier lengths $\ell$ at $\Ray=10^8$ and $\phi=\pi/4$. In $(a)$ the black arrow indicates increasing $\ell$. In $(b)$ $\langle u\rangle_t$ vanishes for $\ell=0$ due to the occurrence of flow reversals.}
\end{figure}

We plot in Fig.~\ref{mean_flow_slice} the profiles of time-averaged temperature $\langle T\rangle_t$ and horizontal velocity $\langle u\rangle_t$ at the horizontal mid-plane for different barrier lengths $\ell$ at $\Ray=10^8$ and $\phi=\pi/4$, to show the effects of the barrier on the mean flow.
Note that for the case of $\Ray=10^8$ and $\ell=0$ the flow exhibits flow reversals, and $\langle u\rangle_t$ vanishes at the horizontal mid-plane.
The temperature and velocity profiles are anti-symmetric with respect to the center point for all values of $\ell$ due to the symmetry of the system under rotation by $\pi$.
Without barrier, the time-averaged temperature in the bulk is uniform and equal to the arithmetic mean temperature $T_m=0.5$, as shown in Fig.~\ref{mean_flow_slice}$(a)$. For $\ell>0$ the temperature values in the interiors of the two sub-regions stay uniform and gradually deviate from $T_m$ for increasing $\ell$, indicating the emergence of a thermal boundary layer on the barrier surface. The overshoots of the temperature profile near the horizontal plates are associated with the formation of the horizontal jets near the barrier ends.
Meanwhile, velocity boundary layers also emerge on the barrier surface with the increase of $\ell$, as shown in Fig.~\ref{mean_flow_slice}$(b)$. The gradients of temperature and velocity on the barrier increase with $\ell$.

\begin{figure}
	\centering
	\includegraphics[width=0.45\linewidth]{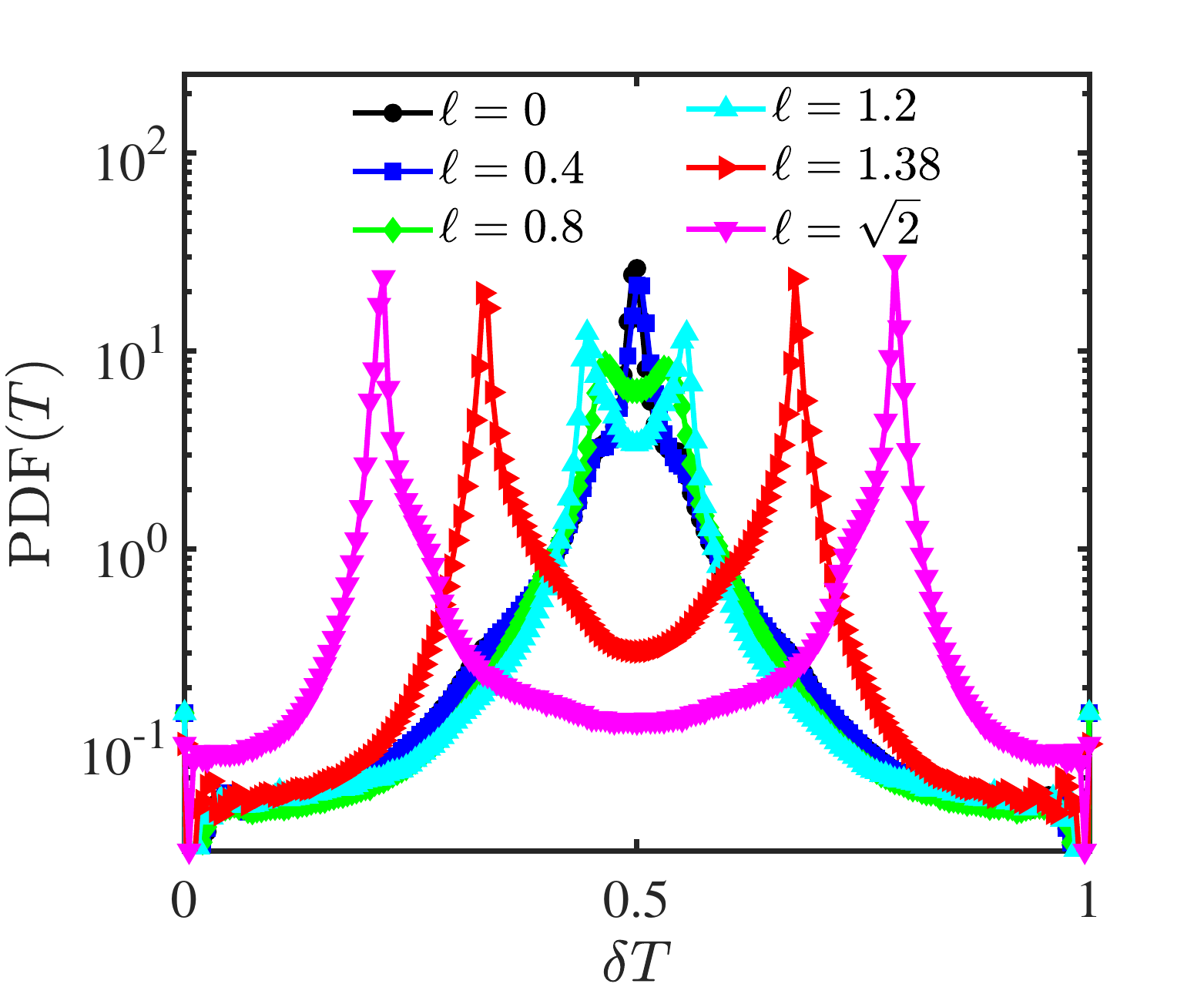}
	\put(-224,170){$(a)$}
	\includegraphics[width=0.45\linewidth]{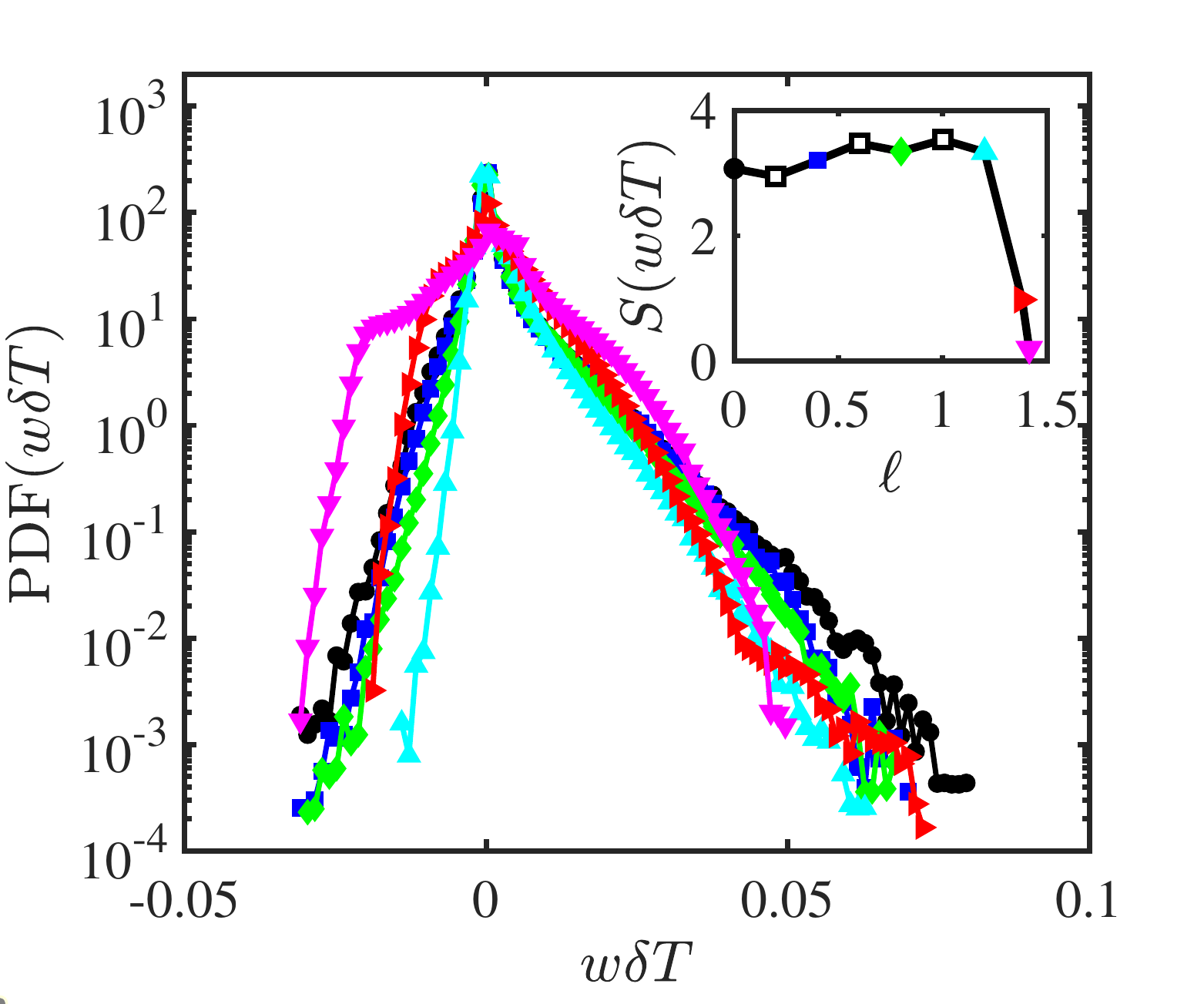}
	\put(-224,170){$(b)$}
	\vspace{-2 mm}
	\caption{\label{pdf_temp_flux}$(a)$ Probability density function $\text{PDF}(\delta T)$ of the temperature fluctuation $\delta T$ for various $\ell$ at $\Ray=10^8$ and $\phi=\pi/4$. $(b)$ $\text{PDF}(w\delta T)$ of the convective heat flux $w\delta T$ in the vertical direction for various $\ell$ at $\Ray=10^8$ and $\phi=\pi/4$. Inset: corresponding skewness $S(w\delta T)$ as a function of $\ell$.}
\end{figure}

We now quantitatively investigate the influence of the barrier on the temperature distribution. Fig.~\ref{pdf_temp_flux}$(a)$ shows the PDFs (probability density functions) of the temperature fluctuation $\delta T=T-T_m$ in the whole cell for various $\ell$ at $\Ray=10^8$ and $\phi=\pi/4$. For different values of $\ell$, $\text{PDF}(\delta T)$ takes a symmetric form, which is consistent with the rotational symmetry of the system. In the traditional RB convection with $\ell=0$, $\text{PDF}(\delta T)$ peaks at $\delta T=0$, which is associated with the well-mixed center core where $T\approx T_m$. When a short barrier is present, the temperature distribution is basically unchanged as in the case of $\ell=0.4$. 
When the barrier becomes longer, the value of $\text{PDF}(\delta T)$ at $\delta T=0$ decreases and $\text{PDF}(\delta T)$ now has two peaks located symmetrically with respect to $\delta T=0$, which is attributed to the reduced temperature mixing in the cell.
For increasing $\ell$ the two peaks become more and more separated, consistent with the time-averaged temperature profiles in Fig.~\ref{mean_flow_slice}$(a)$. We observe that rather than having a single bulk temperature for small $\ell$, we move towards a configuration for which we have an upper and lower bulk temperature for increasing $\ell$. 

How are the local heat transfer properties influenced when symmetry-breaking effects are introduced by a barrier? To address this question, we plot in
Fig.~\ref{pdf_temp_flux}$(b)$ the PDFs of the convective heat flux $w\delta T$ along the vertical direction in the whole cell for $\Ray=10^8$ and $\phi=\pi/4$. The variation of corresponding skewness $S(w\delta T)$ with $\ell$ is shown in the inset.
$\text{PDF}(w\delta T)$ shows a long tail for strong upward convective heat transfer with $w\delta T>0$, resulting from the coupling between the temperature and vertical velocity through the buoyancy term in the momentum equation.
The left branch of $\text{PDF}(w\delta T)$ shows that there is a finite probability for the occurrence of counter-gradient convective heat transfer, as observed in Refs.~\cite{huang2013counter}.
The asymmetric form of $\text{PDF}(w\delta T)$ about $w\delta T=0$ implies that, on average, heat is transferred from the hot bottom plate to the cold top plate.
With the increase of $\ell$, the strength of the counter-gradient convective heat transfer is reduced, as revealed by the decreased probability of the left branch of $\text{PDF}(w\delta T)$. It indicates that due to the symmetry-breaking effects introduced by the barrier, the flow motion is guided---and funneled---in such a way that hotter (colder) fluid tends to flow upwards (downwards), which is beneficial for the heat transfer.
This effect is confirmed by the increase of the skewness $S(w\delta T)$ with $\ell$ as depicted in the insect of Fig.~\ref{pdf_temp_flux}$(b)$, showing that $\text{PDF}(w\delta T)$ becomes more asymmetric for larger $\ell$.
When $\ell$ is further increased to reach $\ell_m$, the probability of the left branch of $\text{PDF}(w\delta T)$ increases due to the enhanced counter-gradient convective heat transfer along the barrier surface, and concomitantly, $S(w\delta T)$ decreases rapidly.

\subsection{Flow reversals}\label{sec:reversal}

In this section we focus on how the barrier influences the flow reversals. To this end we fix the Rayleigh number to $\Ray=10^8$. At this $\Ray$, frequent flow reversals are observed in the traditional RB convection.
An example of the reversal process is shown in movie S1 of the Supplemental Material.
In a square cell, the occurrence of flow reversals can be precisely inferred from the evolution of the global angular momentum of the flow field \cite{sugiyama2010flow,wang2018flow}. The angular momentum with respect to the cell center $o$ (see Fig.~\ref{fig:setup}) is defined as
\begin{equation}
L_o(t)=\left\langle -(z-z_o)u(\vec{x},t)+(x-x_o)w(\vec{x},t)\right\rangle_V,
\end{equation}
where $\langle\hdots\rangle_V$ represents averaging over the entire cell. Positive (negative) values of $L_o$ indicate that the flow field rotates counter-clockwise (clockwise) in the average sense.

\begin{figure} 
	\centering
	\vspace{1 mm}
	\hspace{-2 mm}
	\includegraphics[width=0.51\linewidth]{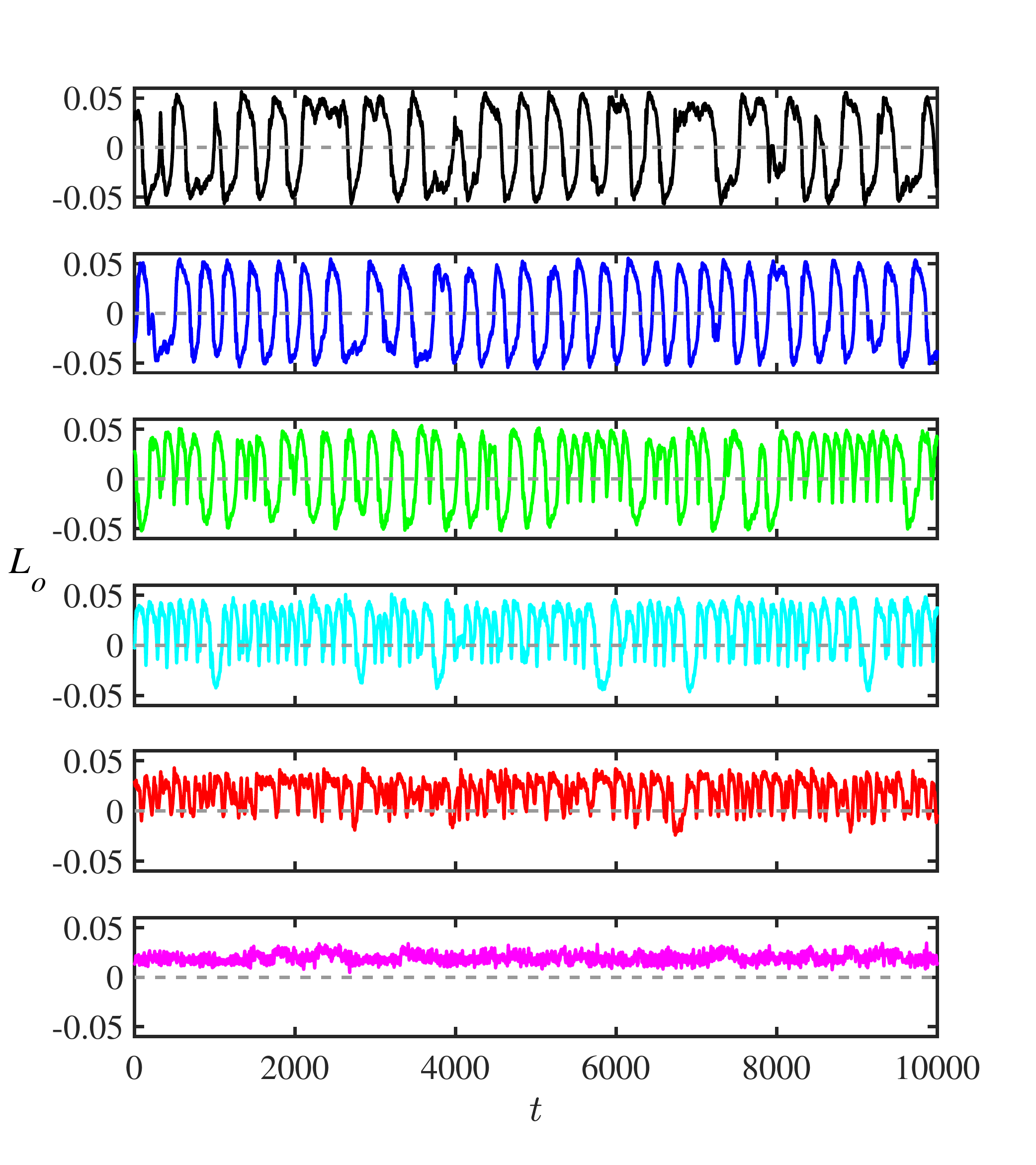}
	\put(-260,265){$(a)$}
	\hspace{-6 mm}
	\includegraphics[width=0.51\linewidth]{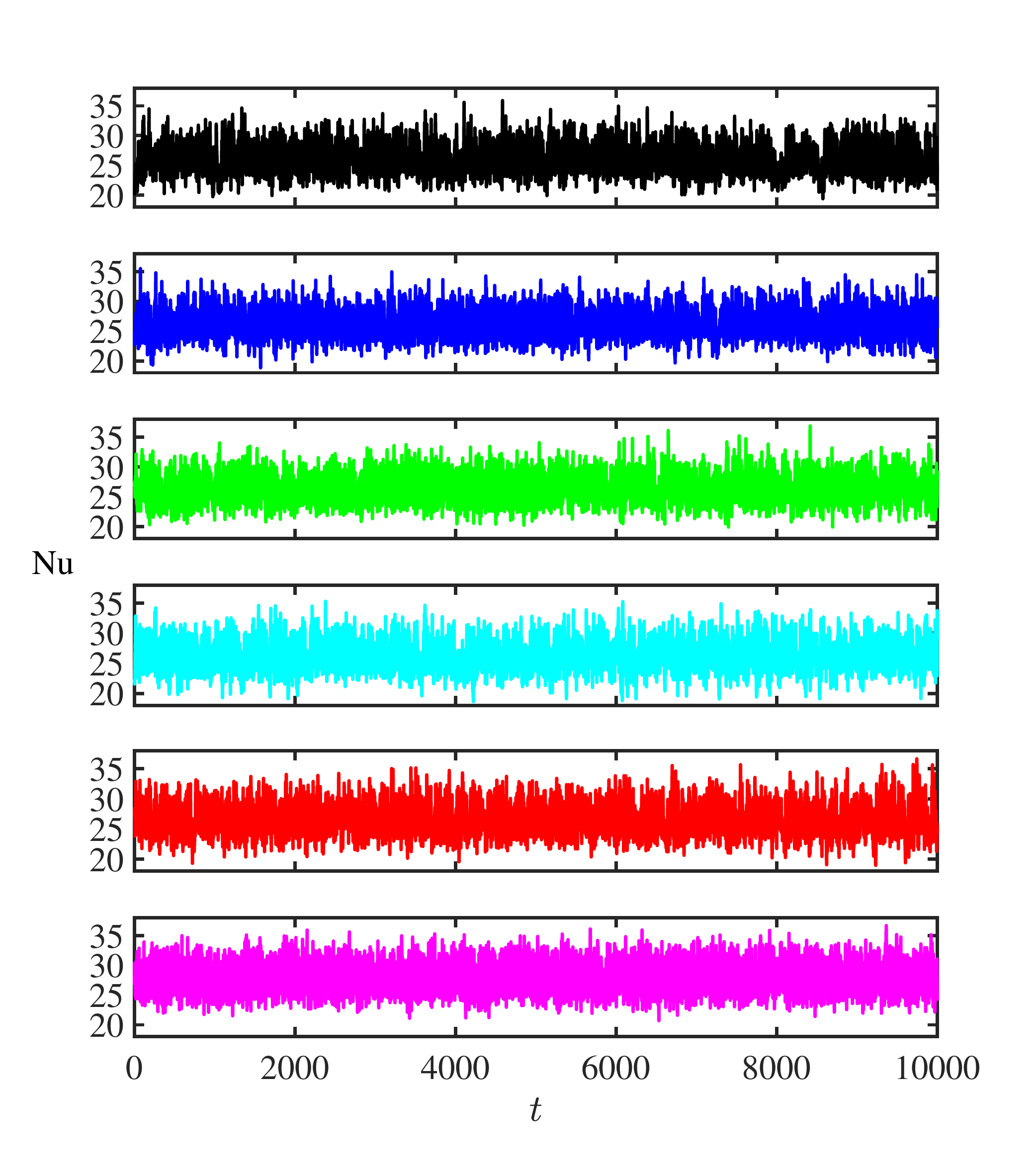}
	\put(-252,265){$(b)$}
	\vspace{-4 mm}
	\caption{\label{time_series} $(a)$ Time series of the angular momentum $L_o(t)$ and $(b)$ Nusselt number $\Nus(t)$ through the top plate for different barrier lengths $\ell$ (indicated by different colors) at $\Ray=10^8$ and $\phi=\pi/4$. In both $(a)$ and $(b)$, the barrier lengths from top to bottom are 0, 0.2, 0.3, 0.4, 0.6, and 0.8.}
\end{figure}

\begin{figure} 
	\centering
	\vspace{1 mm}
	\includegraphics[width=0.45\linewidth]{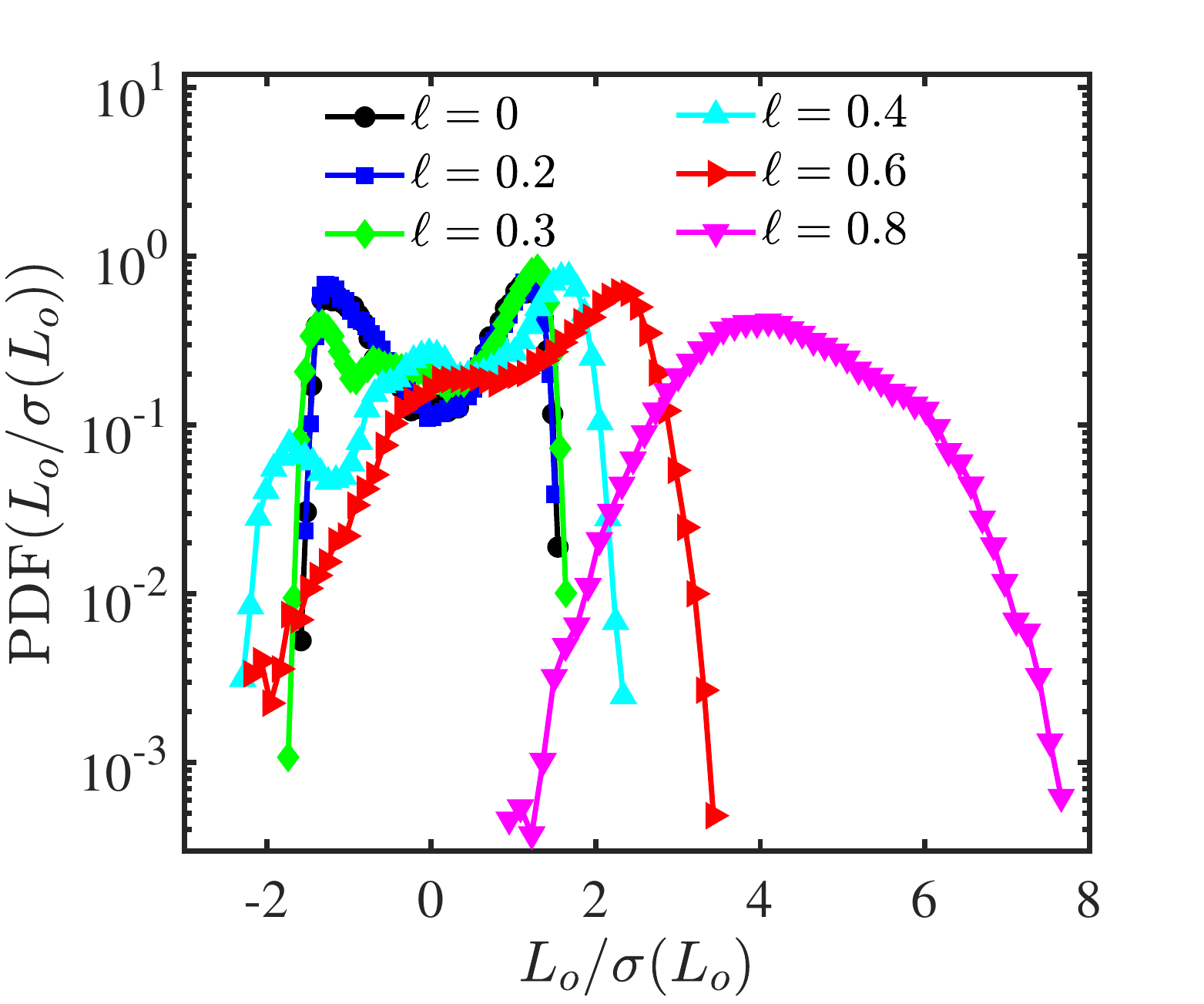}
	\put(-225,170){$(a)$}
	\includegraphics[width=0.45\linewidth]{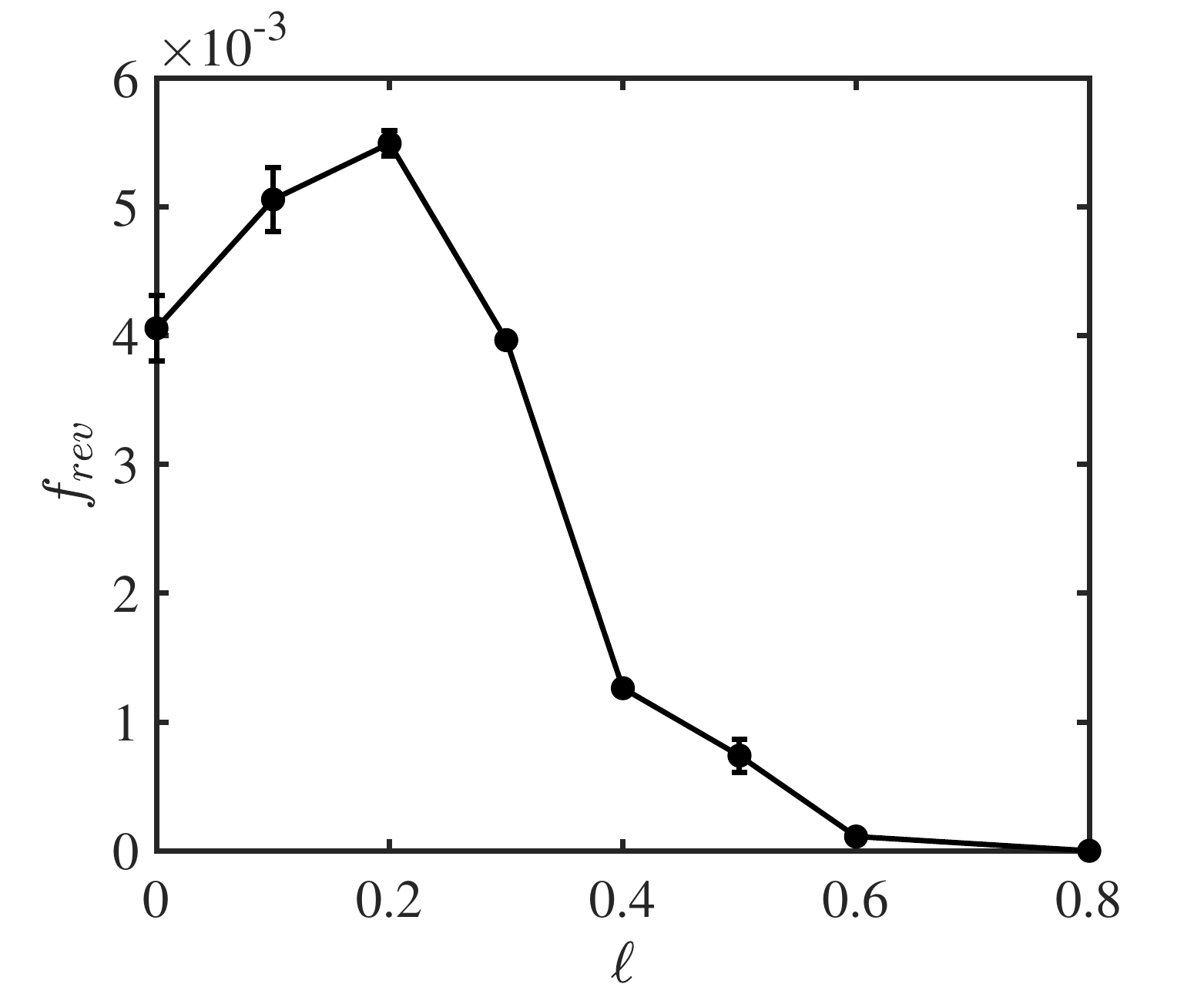}
	\put(-225,170){$(b)$}
	\\
	\includegraphics[width=0.45\linewidth]{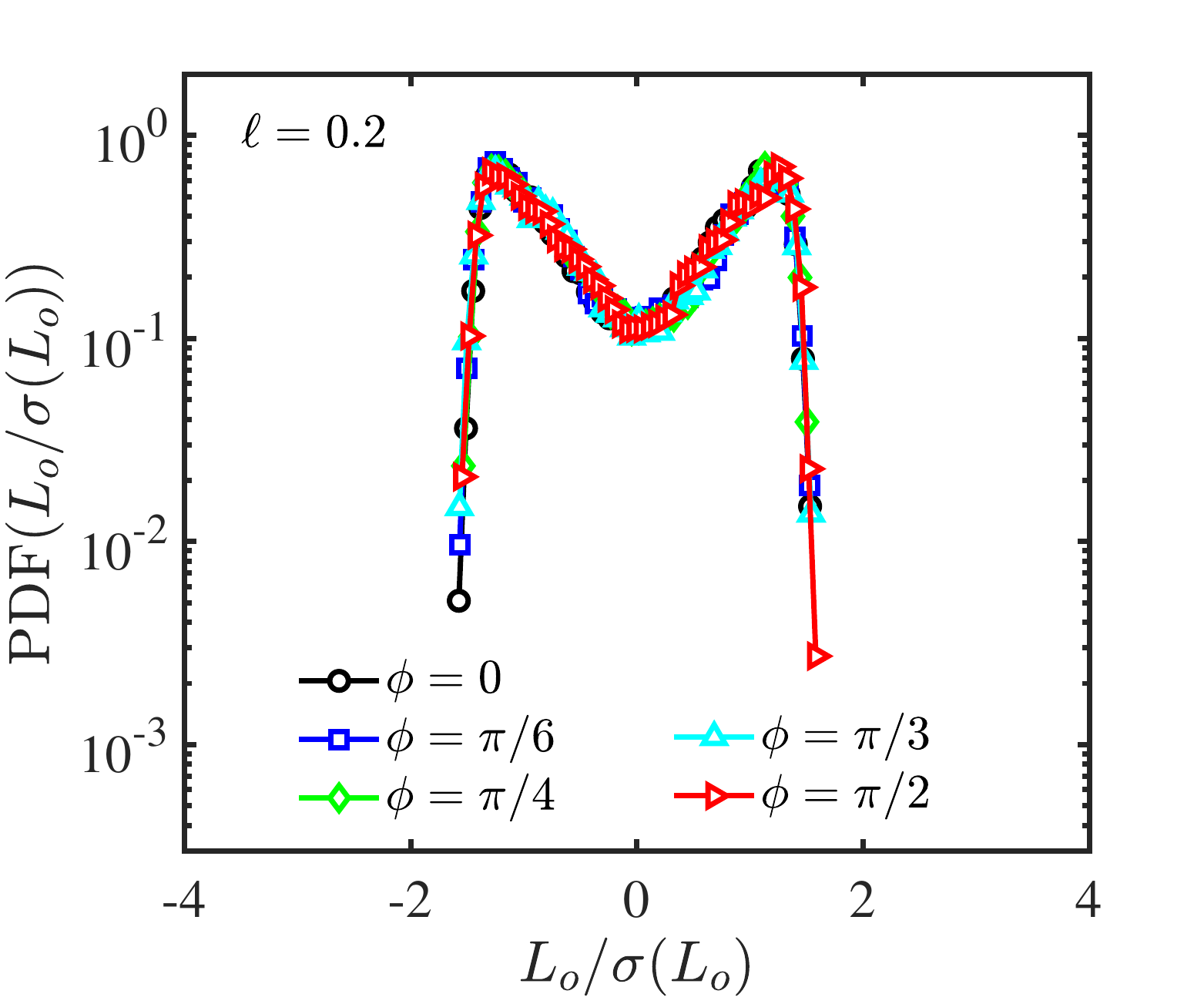}
	\put(-225,170){$(c)$}
	\includegraphics[width=0.45\linewidth]{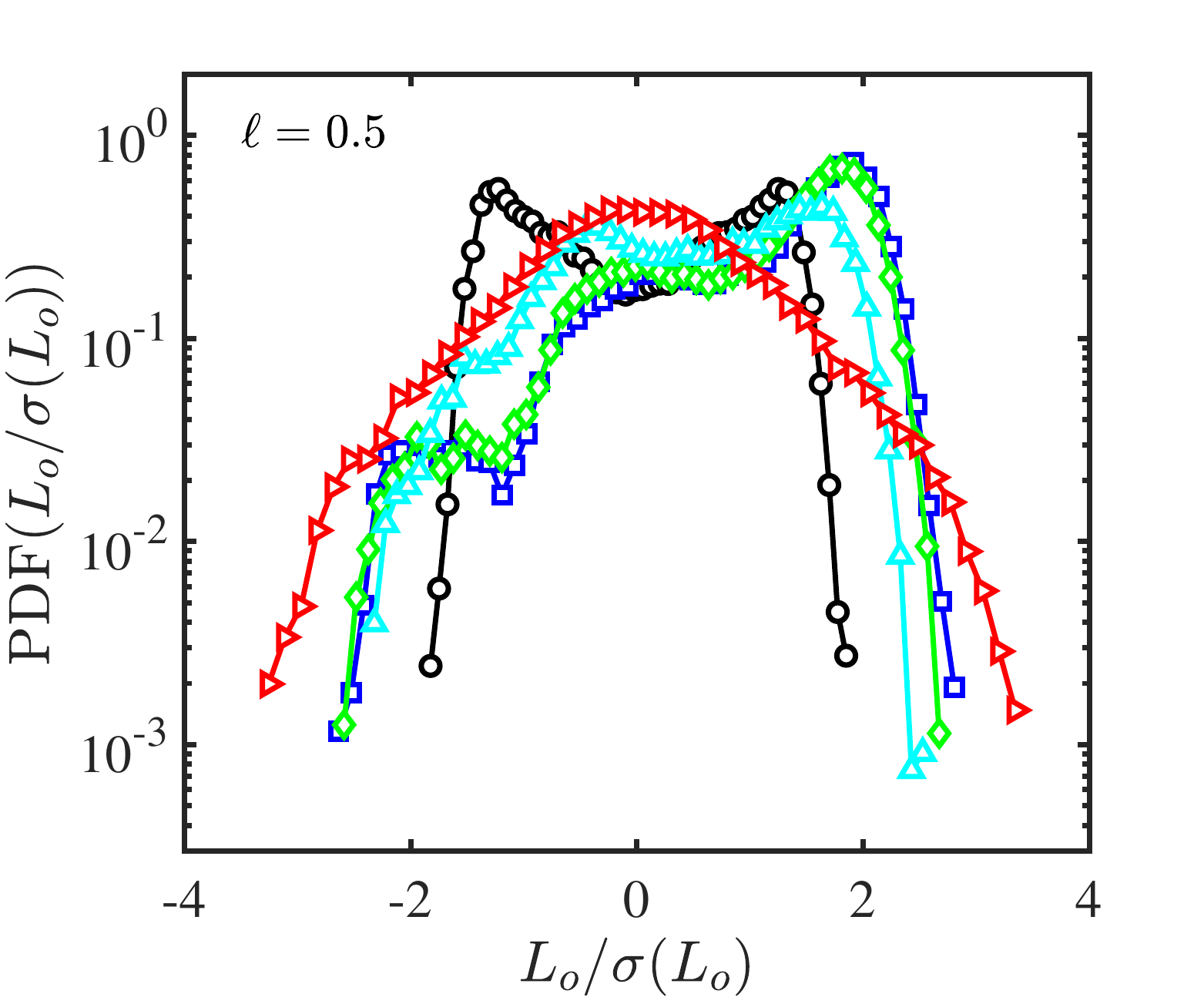}
	\put(-225,170){$(d)$}
	\vspace{-2 mm}
	\caption{\label{reversal} $(a)$ PDFs of the normalized angular momentum $L_o/\sigma(L_o)$ of the time series of $L_o$ in Fig.~\ref{time_series}$(a)$, where $\sigma(L_o)$ is the standard deviation of $L_o$. $(b)$ Reversal frequency $f_{rev}$ as a function of barrier length $\ell$ for $\Ray=10^8$ and $\phi=\pi/4$. The error bar indicates the standard deviation of reversal frequencies based on the first and second halves of the simulations. $(c,d)$ PDFs of $L_o/\sigma(L_o)$ for various $\phi$ and $\Ray=10^8$ with $(c)$ $\ell=0.2$ and $(d)$ $\ell=0.5$.}
\end{figure}

We first examine how the barrier length $\ell$ affects the reversals.
Fig.~\ref{time_series}$(a)$ shows the time series of $L_o$ for different $\ell$ at $\phi=\pi/4$. Corresponding PDFs of $L_o$ are given in Fig.~\ref{reversal}$(a)$. 
The statistical samples are large enough to have good statistical convergence. The longest simulation was run for around $8\times 10^4$ free-fall times.
In the absence of a barrier, frequent reversal events are observed, and $\text{PDF}(L_o)$ has a symmetric, bi-modal shape, demonstrating that the flow states with clockwise and anti-clockwise LSCs are equally probable. These two flow states are equivalent due to the parity symmetry of the system with respect to the horizontal mid-plane. Transition between the two flow states occurs randomly.
The reversal frequency $f_{rev}$ is given in Fig.~\ref{reversal}$(b)$. The reversal events were detected based on the time series of $L_o(t)$. When $L_o\ge\sigma_p(L_o)$ ($L_o\le-\sigma_p(L_o)$), the flow is deemed to be in a well-developed anti-clockwise (clockwise) state, where $\sigma_p(L_o)$ is the standard deviation of $L_o$ with $L_o>0$. In doing so, it is assumed that the typical values of $|L_o|$ of the fully-developed anti-clockwise and clockwise states are comparable for the $\ell$ values considered in Fig.~\ref{reversal}$(b)$ ($\ell\le0.8$). Except for the three cases with largest $\ell$, at least 60 reversal events were detected to calculate $f_{rev}$ (and for some cases we detected even 300 reversal events). We like to highlight that it is extremely difficult to obtain sufficient samples of reversal events for large $\ell$ as reversals become rare, as shown in Fig.~\ref{reversal}$(b)$.

When a short barrier is included in the cell, the parity symmetry of the original system is partially broken. As shown in Fig.~\ref{flow_structure}, the flow field still consists of a well-developed LSC and two corner rolls. However, the states with counter-rotating LSCs are no longer equivalent. Despite that, frequent reversals between the two states are observed for $\ell=0.2$, and $\text{PDF}(L_o)$ still takes a symmetric, bi-modal profile in Fig.~\ref{reversal}$(a)$. Thus, a very short barrier does not qualitatively influence the reversal process. Quantitatively, the reversal frequency is slightly increased, as shown in Fig.~\ref{reversal}$(b)$. It is attributed to the fact that a short barrier can distort and dampen the LSC and affect the competition between the LSC and corner rolls.

With the increase of $\ell$, the symmetry-breaking effects of the barrier become more significant, strongly influencing the reversal events. The time series of $L_o$ for $\ell=0.3$ and 0.4 in Fig.~\ref{time_series}$(a)$ show that reversal events still exist. However, it is found that the negative values of $L_o$ become less probable as compared to the positive values, which is also demonstrated by the asymmetric profiles of $\text{PDF}(L_o)$ in Fig.~\ref{reversal}$(a)$. The asymmetry of $\text{PDF}(L_o)$ originates from the fact that there is no complete reversal following some of the cessations of the anti-clockwise state. When a cessation occurs, $L_o$ decreases and vanishes. Then $L_o$ becomes negative due to the inertia. There is a finite probability for $L_o$ to become again positive after being negative transiently, indicating the re-establishment of the anti-clockwise state. 
See movie S2 in the Supplemental Material for an illustration of the unsuccessful reversal process.
Due to the strong symmetry-breaking effects, the flow prefers the anti-clockwise state statistically.
$\text{PDF}(L_o)$ displays three bumps at $\ell=0.3$ and 0.4 in Fig.~\ref{reversal}$(a)$. The right and left bumps correspond to the anti-clockwise and clockwise states, respectively, and the middle one is associated with the frequent cessation events without complete reversal. 
For even larger $\ell$, the cessation events are also inhibited, and the rotation of the flow is biased in one direction, as illustrated by movie S3 in the Supplemental Material.
Correspondingly, $f_{rev}$ gradually decreases and eventually vanishes for large enough $\ell$, as shown in Fig.~\ref{reversal}$(b)$. It is expected that there exists a critical barrier length above which there is no reversals. The reason for the lack of reversals is possibly due to two mechanisms: 1.~The normal process for a reversal is the gradual growth of the corner rolls after which it has gained enough angular momentum to overturn the entire flow. For the case of $\phi \approx \pi/4$, the barrier is actively limiting the size of the corner rolls, and thus prevents enough angular momentum build-up in the corner rolls to overturn the flow. 2.~For large barrier length the funnelling of the flow is actively resisting the reversal of the flow as the cold fluid would like to go down and the hot up, the least resistance would be to use the funneling mechanism of the barrier rather than moving horizontally.

We show in Figs.~\ref{reversal}$(c,d)$ the PDFs of $L_o$ for various $\phi$ at $\ell=0.2$ and $\ell=0.5$, respectively. When the barrier is very short, the profiles of $\text{PDF}(L_o)$ for different barrier angles collapse, indicating that the flow dynamics are qualitatively unaffected by the barrier. 
$\text{PDF}(L_o)$ takes a symmetric, bi-modal form, demonstrating that the LSC can take the clockwise and anti-clockwise directions with similar probabilities, and flow reversal occurs, even though the parity symmetry of the system is slightly broken with $0<\phi<\pi/2$.
When the barrier becomes longer, the effects of the barrier on the flow dynamics become stronger, as revealed by the profiles of $\text{PDF}(L_o)$ for various $\phi$ at $\ell=0.5$ in Fig.~\ref{reversal}$(d)$. When the barrier aligns horizontally with $\phi=0$, $\text{PDF}(L_o)$ takes a symmetric, bi-modal form. The symmetry of $\text{PDF}(L_o)$ is a manifestation of the intrinsic symmetry of the system. The bi-modal form indicates the occurrence of reversal events. When $0<\phi<\pi/2$, $\text{PDF}(L_o)$ becomes asymmetric due to the strong symmetry-breaking effects introduced by the barrier. When $\phi=\pi/2$, $\text{PDF}(L_o)$ recovers to a symmetric form, consistent with the symmetry of the system. In contrast with the $\phi=0$ case, $\text{PDF}(L_o)$ at $\phi=\pi/2$ has only one peak located at $L_o=0$. The discrepancy is associated with the different flow organizations. For $\phi=\pi/2$ both the left and right sub-regions prefer a vertically-stacked two-roll state, rather than having a single large global roll. Thus, $L_o$ has a high probability to vanish. While for $\phi=0$ the simulation shows that both the top and bottom sub-regions prefer a single-roll state. Thus, $L_o$ generally takes a negative or positive value depending on the orientation of the rolls.

The time-series of $\text{Nu}$ are shown in Fig.~\ref{time_series}$(b)$, corresponding to the cases in Fig.~\ref{time_series}$(a)$. The temporal dynamics of the Nusselt number have much higher frequencies (related to plume detachment and plume impact) as compared to the dynamics of the angular momentum (related to the reversals of the flow). It is hard to find any signature of reversals---clearly seen in Fig.~\ref{time_series}$(a)$---in $\Nus(t)$.

\FloatBarrier

\section{Conclusions}\label{sec:summary}
In summary, we performed a numerical study of turbulent RB convection with a single, centered, passive, conductive barrier.
The barrier is added aiming to control the LSC, the thermal transport, and to funnel the hot and cold plumes.
The influences of the barrier on the heat transfer, flow organization, and reversal events are studied for a myriad of barrier lengths $\ell$ and angles $\phi$.
In the presence of a short barrier, the heat transfer and flow structure are only mildly affected. With the increase of $\ell$, the influences of the barrier become stronger. We find that the global heat transfer can be considerably enhanced (up to 18\%) even though the average flow velocity is significantly reduced, which is due to the guiding and focusing of the cold and hot plumes. The local flux on the horizontal plates is also greatly enhanced (up to 800\% for the parameters considered).
The flow organization can be significantly changed due to the constraints of a long barrier. With the increase of $\ell$, thermal and kinetic boundary layers gradually develop on the barrier surface, and the gradients of temperature and velocity on the barrier are dependent on the length and angle of the barrier.
When the gap between the barrier end and the cell boundaries is small, strong jets form across the gap due to the convergence of fluid flow and the pressure imbalance. The interaction between the jets and boundary layers is beneficial for the heat transfer due to enhanced plume generation.
When the barrier is long enough, the convection cell is divided into two sub-regions without mass transfer between each other. In this case, the heat transfer and flow dynamics are highly dependent on the barrier angle $\phi$. For large $\phi$, each of the two sub-regions contains part of the top and bottom plates, and the global heat transfer is mildly influenced by the barrier; while when $\phi$ is small, the system can be regarded as a combination of two vertically-stacked RB cells with reduced effective $\Ray$, and the heat transfer is significantly reduced.

To further examine the effects of the barrier on the LSC, the influence of the barrier on the flow reversal is studied. When the barrier is very short, the reversal process is qualitatively unchanged for different values of $\phi$, even though the parity symmetry of the system is slightly broken for $0<\phi<\pi/2$. When the barrier becomes longer, flow reversals are gradually suppressed due to the stronger symmetry-breaking effect of the barrier, which demonstrates the stabilizing effects of the barrier on the LSC.

The results highlight that the turbulent thermal transport can be enhanced by a passive barrier even though the flow strength is reduced, demonstrating a simple and promising way to enhance the heat transfer in technological applications. In the future, it will be of interest to extend this study to larger parameter space, including $\Pra$, $\Gamma$, and multiple barriers to achieve even better flow control.

\begin{acknowledgments}
We thank Chao Sun for feedback on and discussion of our results. This work was financially supported by the Natural Science Foundation of China under grant 11988102, 91852202, and 11672156 and by the Zwaartekracht programme Multiscale Catalytic Energy Conversion (MCEC), which is part of The Netherlands Organisation for Scientific Research (NWO). S.L. acknowledges the project funded by the China Postdoctoral Science Foundation under Grant No. 2019M660614.
\end{acknowledgments}

\FloatBarrier

\appendix

\section{}\label{appA}

Numerical details of the typical grid resolutions in the 2D simulations with $\phi=\pi/4$ are provided in Table \ref{tab:resolution}. A finer resolution is used when $\ell$ is close to $\ell_m$ as a `jet' is created between the horizontal plate and the barrier end and the thermal boundary layer near the barrier end becomes thinner.
\begin{table}[htbp!]
	\centering
	\renewcommand\arraystretch{1.2}
	\setlength{\tabcolsep}{3mm}
	\begin{tabular}{ccccc}
		\hline
		$\ell$       & $\Ray=10^6$      & $\Ray=10^7$      & $\Ray=10^8$      & $\Ray=10^9$      \\ \hline
		$\ell<1.2$ & 180$\times$180 & 180$\times$180 & 280$\times$280 & \textbf{540$\times$540} \\
		$\ell=1.2$  & 180$\times$180 & 180$\times$180 & 280$\times$280 & 900$\times$540 \\ 
		$\ell=1.3$  & 300$\times$300 & 300$\times$300 & 400$\times$400 & 1080$\times$540 \\ 
		$\ell=1.35$  & 400$\times$480 & 450$\times$480 & 1080$\times$480 & 2250$\times$540 \\
		$\ell=1.38$  & -- & -- & 2000$\times$864 & -- \\
		$\ell=\sqrt{2}$  & 400$\times$400 & 400$\times$400 & 540$\times$540 & 800$\times$800 \\		\hline
	\end{tabular}
	\caption{\label{tab:resolution} Typical grid resolutions $N_z\times N_x$ for different $\Ray$ and $\ell$ in the 2D simulations with $\phi=\pi/4$. The grid is uniform in the horizontal direction and stretched in the vertical direction with grid points clustered near the horizontal plates. For the bold case we show for the case of $\ell=1.0$, the grid and the ratio of the grid size and the local Kolmogorov length scale in Fig.~\ref{fig:grid}.}
\end{table}

\begin{figure}
	    \centering
	    \hspace{-3 mm}
	    \includegraphics[width=0.25\linewidth]{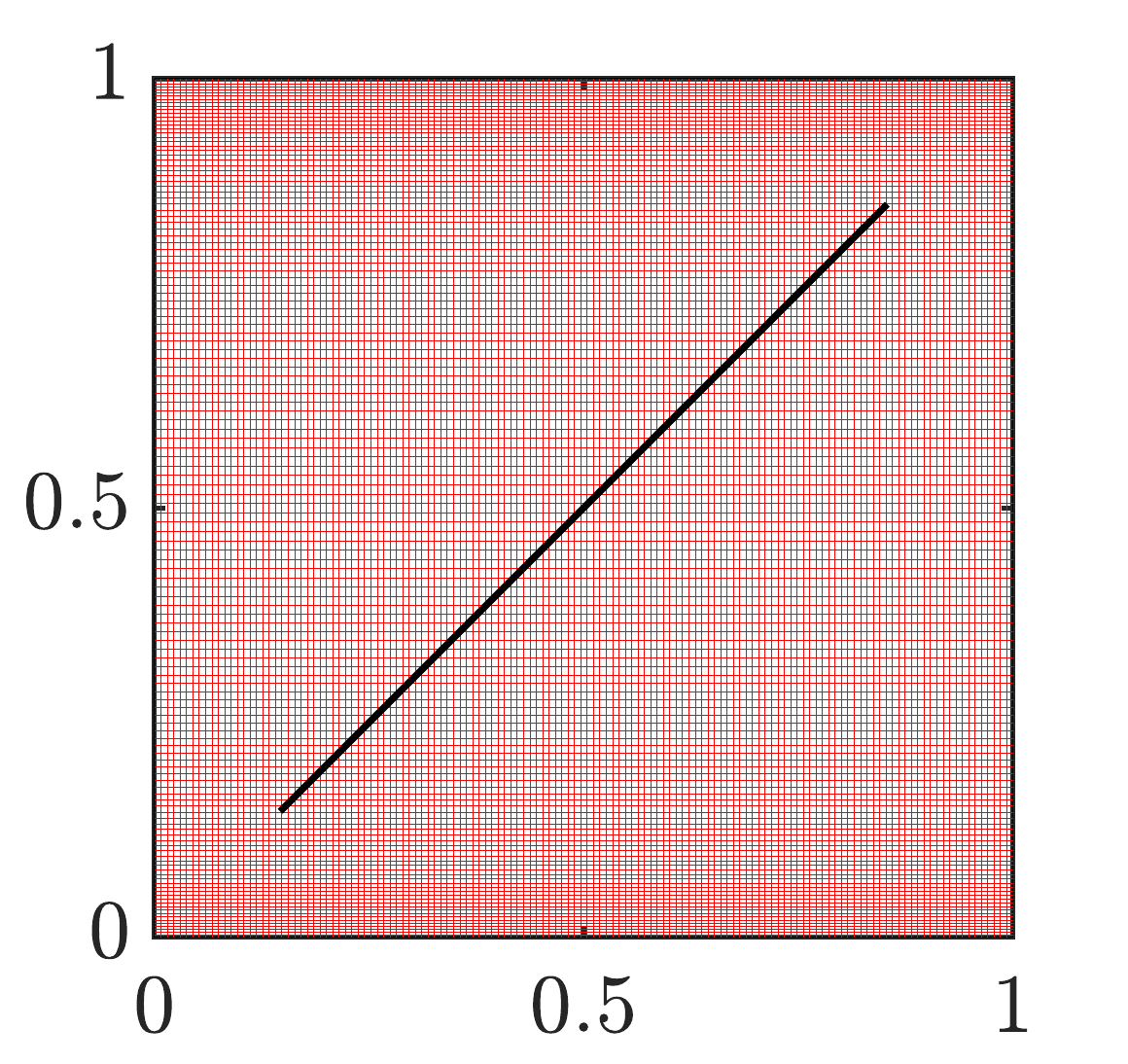}
	    \put(-135,105){$(a)$}
	    \hspace{4 mm}
	    \includegraphics[width=0.25\linewidth]{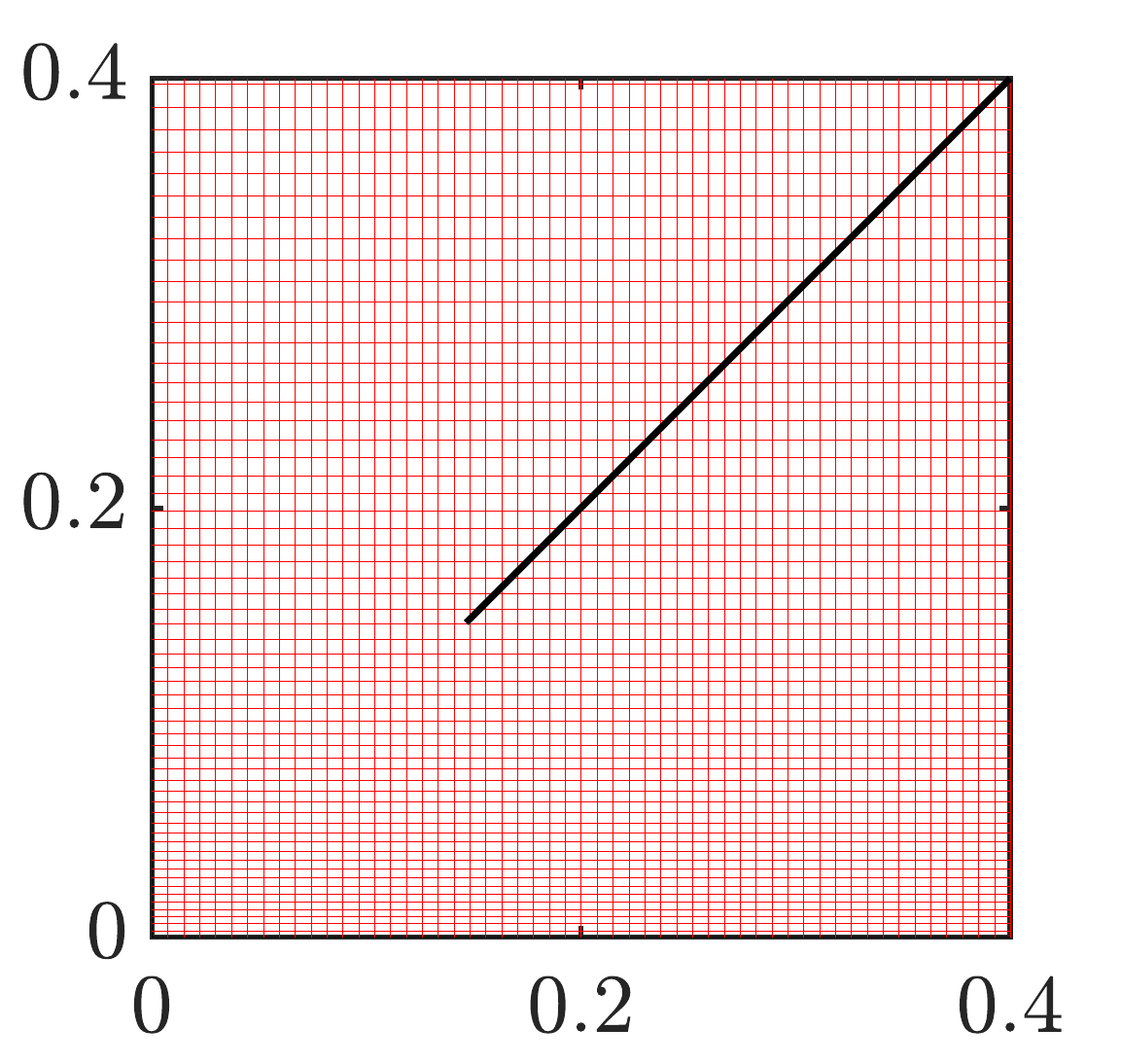}
	    \put(-138,105){$(b)$}
	    \hspace{-2 mm}
	    \includegraphics[width=0.302\linewidth]{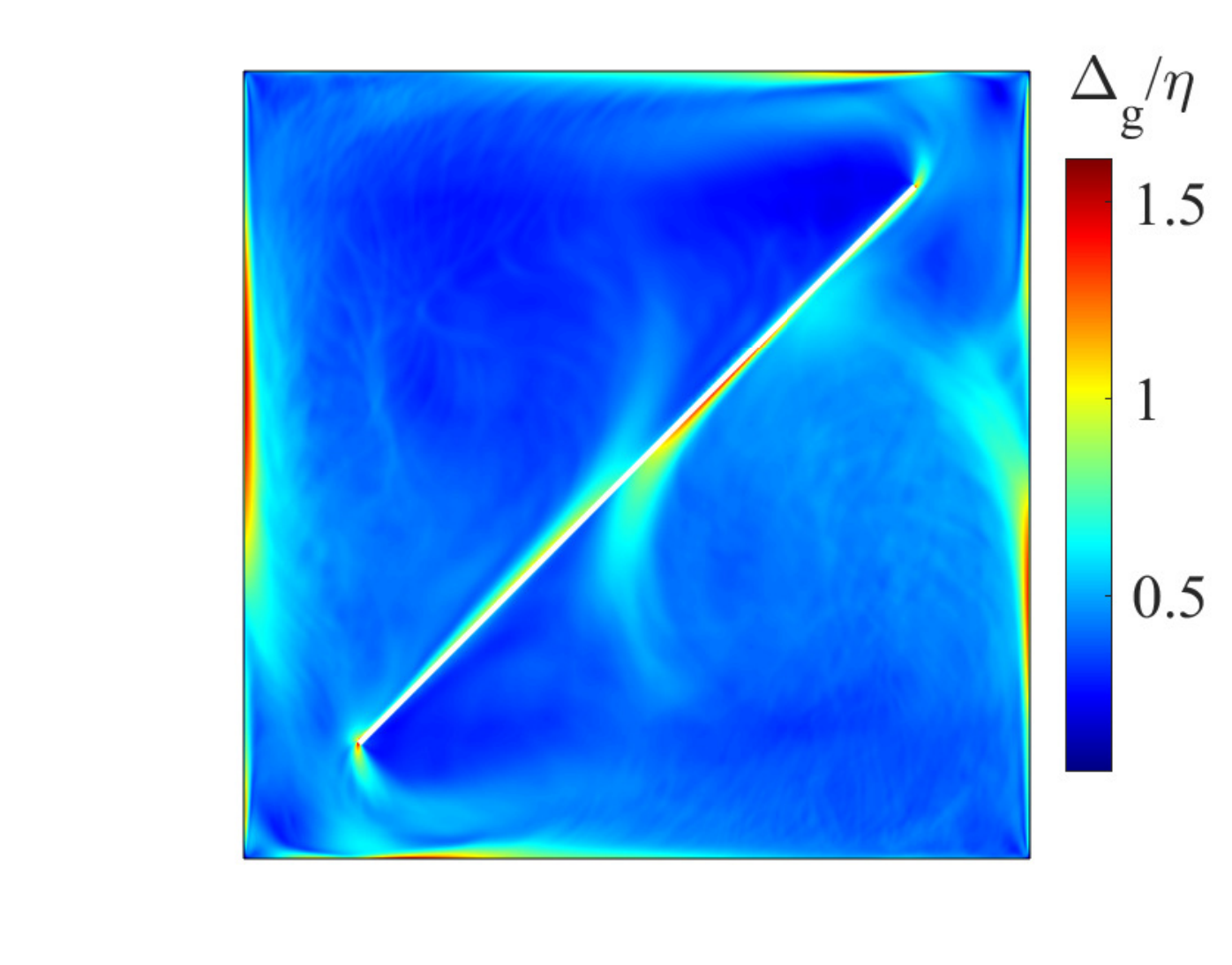}
	    \put(-140,105){$(c)$}
	    \caption{$(a)$ An example of the grid used in the simulations with $N_z\times N_x=540\times540$. \textbf{The grid is displayed every fourth grid lines in both the horizontal and vertical directions.} The black solid line indicates a barrier with $\ell=1.0$ and $\phi=\pi/4$.
	    An enlargement for $(x,z)\in[0,0.4]\times[0,0.4]$ is shown in $(b)$. $(c)$ The ratio between the local grid spacing $\Delta_g$ and the local Kolmogorov length scale $\eta$ at $Ra=10^9$, $\ell=1.0$, $\phi=\pi/4$. The maximum ratio is 1.61.}
	    \label{fig:grid}
	\end{figure}

    An example of a grid with $N_z\times N_x=540\times540$ is shown in Figs.~\ref{fig:grid}$(a,b)$.
    The vertical grid spacing in the boundary layer region is slightly denser than that in the central region.
    Fig.~\ref{fig:grid}$(c)$ shows the ratio of the local grid size $\Delta_g=\max(\Delta z,\Delta x)$ ($\Delta x$  is the horizontal dimension of the grid and $\Delta z$ the vertical dimension of the grid) and local Kolmogorov microscale $\eta=(\nu^3/\langle\epsilon_u\rangle_t)^{1/4}$ for $Ra=10^9$ and $\ell=1.0$. We estimate $\eta$ by the time-averaged kinematic energy dissipation rate:
	\begin{align}
	\langle\epsilon_u\rangle_t &= \left\langle \frac{1}{2}\nu\sum_{ij}\left[\frac{\partial u_j}{\partial x_i}+\frac{\partial u_i}{\partial x_j}\right]^2 \right\rangle_t.
	\end{align}
	The maximum value of $\Delta_g/\eta$ is 1.61 which is deemed sufficient \cite{pope2000turbulent}. Note that we have used $\max(\Delta z,\Delta x)$ for the grid spacing which is the worst-case scenario (an overestimate) as in the boundary layers at the top and bottom plates the largest gradients are in the vertical direction and the largest grid size is in the horizontal direction. Sufficient number of nodes are located in the boundary layers, which agrees with the recommendation by Ref.~\cite{shishkina2010boundary}.
	We also estimate the Kolmogorov scale by the global criterion $\eta_g=LPr^{1/2}/[Ra(Nu-1)]^{1/4}$. The grid spacing is found to satisfy $\Delta_g\le 0.63\eta_g$.

To further guarantee the sufficiency of grid resolution, grid independence studies were performed at selected parameters, particularly for large $\ell$, where the barrier end is close to the wall, and thermal and velocity boundary layers develop on the barrier surfaces, as shown in Fig.~\ref{mean_flow_slice}. The numerical details of the grid independence studies are shown in Table \ref{tab:grid_independence}. Overall, we find that the relative errors of $\Nus$ and $\Rey$ based on two grids of distinct resolutions are all within $1\%$ or less.

\begin{table}[htbp!]
	\centering
	\renewcommand\arraystretch{1.2}
	\setlength{\tabcolsep}{3mm}
	\begin{tabular}{ccccccc}
		\hline
		$(\Ray,\ell)$     & $N_z\times N_x$ & $\Nus$  & $\Rey$   & $(\epsilon_{\Nus},\epsilon_{\Rey})$  \\ \hline
		$(10^7,1.2)$  & $180\times 180$ & 15.41 & 89.23  & \multirow{2}{*}{(0.3\%,0.7\%)}  \\ 
		$(10^7,1.2)$  & $400\times 400$ & 15.36 & 88.62  &   \\ \hline
		$(10^7,1.35)$  & $450\times 480$ & 14.32 & 71.42  & \multirow{2}{*}{(0.6\%,0.4\%)}  \\ 
		$(10^7,1.35)$  & $800\times 800$ & 14.23 & 71.16  &   \\ \hline
		$(10^7,\sqrt{2})$  & $400\times 400$ & 7.60 & 79.53  & \multirow{2}{*}{(-0.4\%,0.5\%)}  \\ 
		$(10^7,\sqrt{2})$  & $800\times 800$ & 7.63 & 79.17  &   \\ \hline
		$(10^8,1.2)$  & $280\times 280$ & 26.92 & 276.8  & \multirow{2}{*}{(0.1\%,-0.1\%)}  \\ 
		$(10^8,1.2)$  & $600\times 600$ & 26.88 & 276.9  &   \\ \hline
	    $(10^9,1.0)$ & $540\times 540$   & 57.66 & 1475 & \multirow{2}{*}{(0.2\%,0.3\%)}\\
	    $(10^9,1.0)$ & $1080\times 1080$ & 57.54 & 1471 &  \\ \hline
	\end{tabular}
	\caption{\label{tab:grid_independence} Numerical details of the grid independence studies at selected parameters with $\phi=\pi/4$. From left to right the columns indicate the physical parameters $(\Ray,\ell)$, the grid resolution $N_z\times N_x$ along the vertical and horizontal directions, the Nusselt number $\Nus$, the Reynolds number $\Rey$, and the relative errors $(\epsilon_{\Nus},\epsilon_{\Rey})$ of $(\Nus,\Rey)$ based on two grids of different resolutions. The typical statistical error for $\Nus$ (standard error of the mean) is within 1\%.}
\end{table}

\vspace{-3 mm}
\FloatBarrier
\section{}\label{appB}

\begin{figure}[htbp!]
	\centering
	\includegraphics[width=0.72\linewidth]{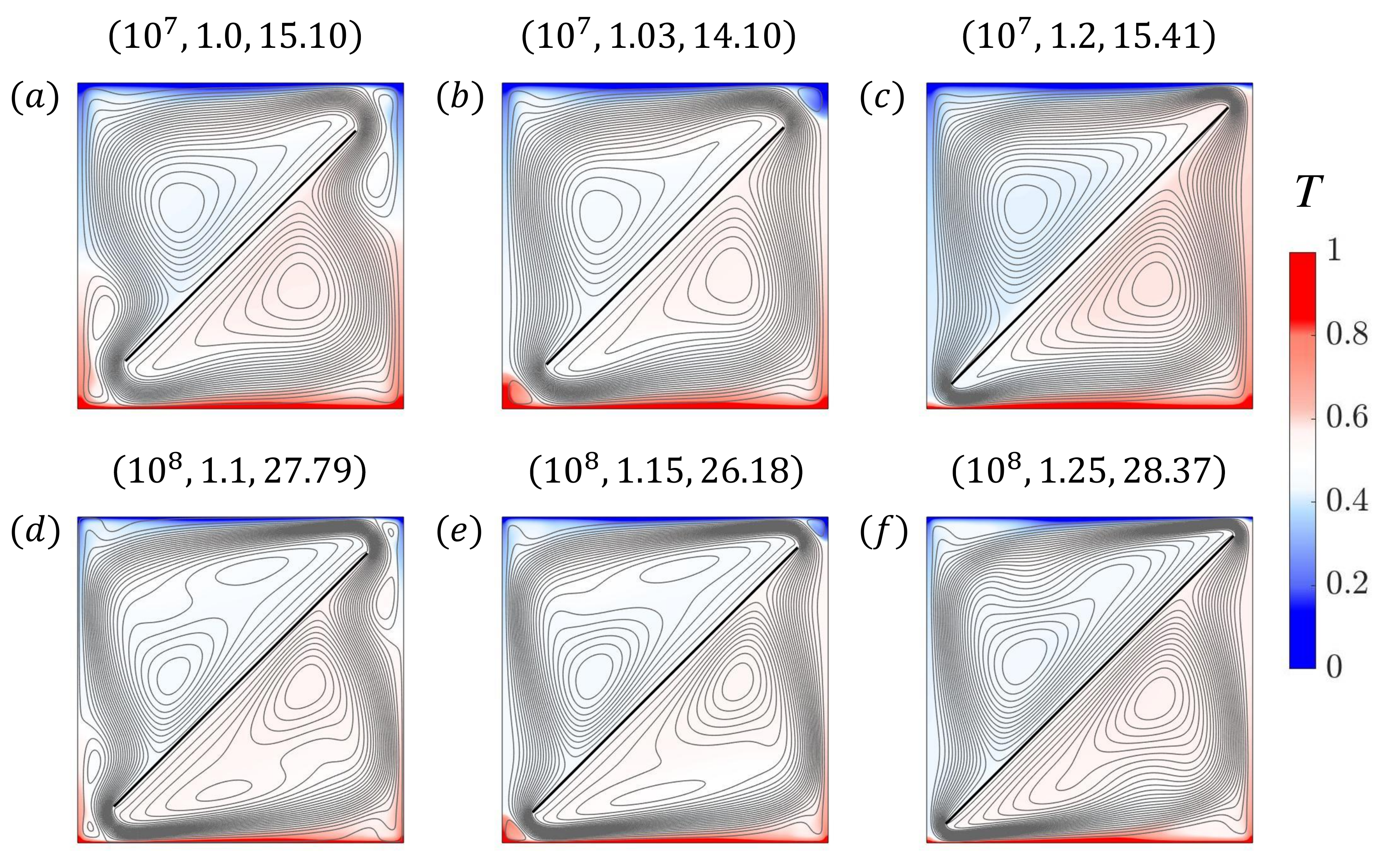}
	\vspace{-2 mm}
	\caption{\label{mean_flow_plume}Streamlines based on the average flow field superimposed on the temperature field for increasing barrier length $\ell$ at $(a-c)$ $\Ray=10^7$ and $(d-f)$ $\Ray=10^8$ for a barrier angle of $\phi=\pi/4$. Plot labels for each panel indicate $(\Ray,\ell,\Nus)$.}
\end{figure}

\begin{figure}[htbp!]
	\centering
	\includegraphics[width=0.42\linewidth]{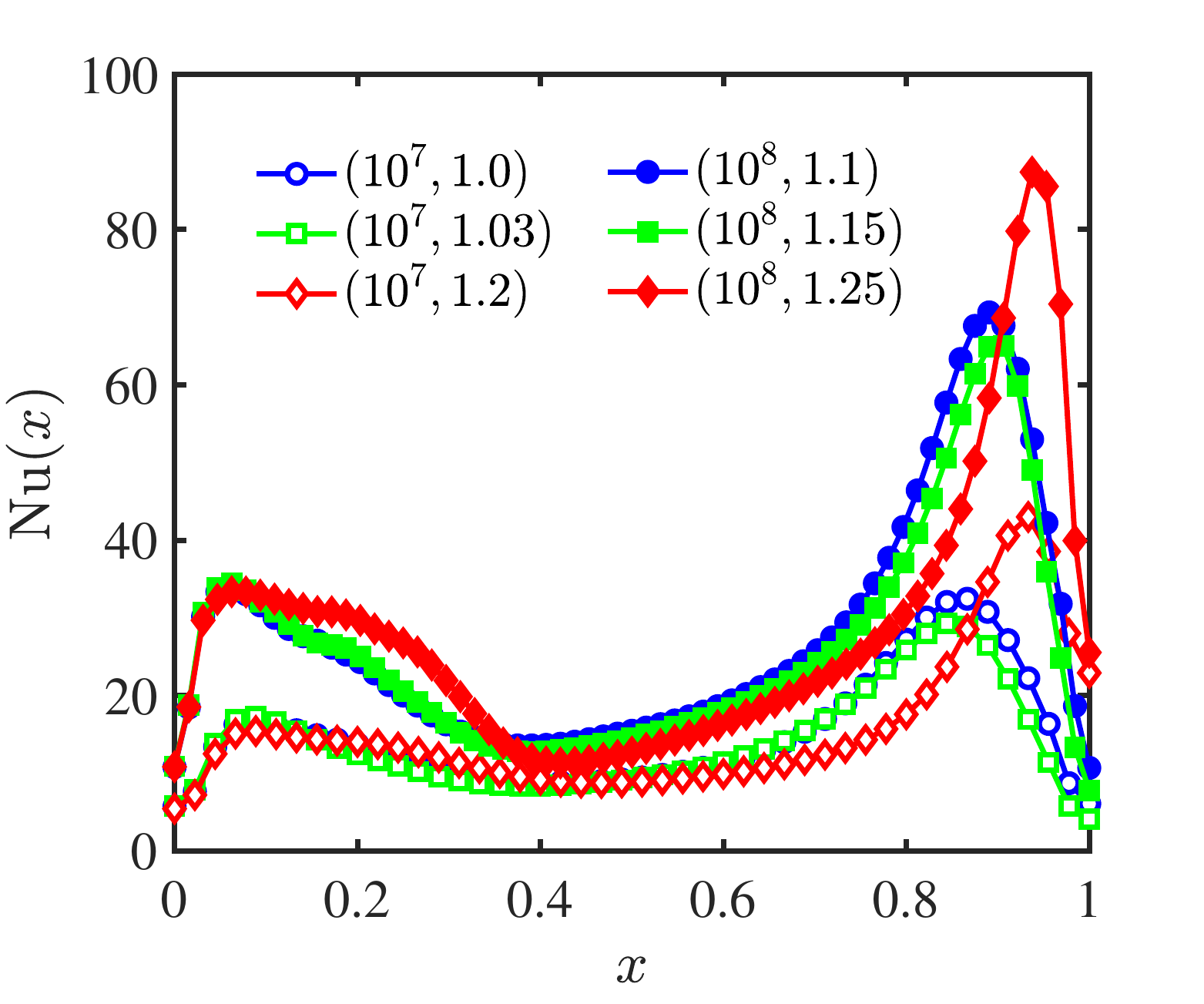}
	\vspace{-2 mm}
	\caption{\label{local_Nu_anomaly}Time-averaged local flux $\Nus(x)$ on the top plate for the same parameters as Fig.~\ref{mean_flow_plume}. The values of $(\Ray,\ell)$ are labeled. All cases are for a barrier angle of $\phi = \pi/4$.}
\end{figure}

For certain parameters $(\ell,\phi)$, the interaction between the barrier and corner flow is strong, and a mild variation of $\ell$ can significantly change the flow structure in the corner and the heat transfer.
Fig.~\ref{mean_flow_plume} shows the time-averaged temperature fields superimposed with streamlines for three values of $\ell$ at $\Ray=10^7$ and $\Ray=10^8$ for a barrier angle of $\phi=\pi/4$. 
Corresponding profiles of the time-averaged local heat flux $\Nus(x)$ on the top plate are displayed in Fig.~\ref{local_Nu_anomaly}, showing the appearance of a peak near the barrier end, as already depicted in Fig.~\ref{local_flux}$(a)$.
For both values of $\Ray$, at relatively small $\ell$, plumes arise from the thermal boundary layers along the sidewalls. In the top-right and bottom-left corners, small circulations are observed, which are connected to separation vortices attached on the sidewalls. When $\ell$ is slightly increased, the size of the plumes in the top-right and bottom-left corners is decreased, and the separation vortices are suppressed.
Furthermore, a marked drop of the global heat transfer is observed with even a slight increase of $\ell$.
From Fig.~\ref{local_Nu_anomaly} it is found that with the increase of $\ell$, the peak value $\Nus_{\text{max}}$ of $\Nus(x)$ is decreased, demonstrating the influence of the change of corner flow structure on the heat transfer.
As $\ell$ is further increased, the circulations in the top-right and bottom-left corners are also suppressed, while $\Nus_{\text{max}}$ is significantly increased, due to the impact of strong flow of hot and cold fluids on the opposite boundary layers. The global $\Nus$ is also significantly increased.
Thus, the change of the local flow structure in the corner and near the boundary layer has a significant influence on the heat transfer, and a gradual variation of $\ell$ may result in an evident and non-monotonic change of $\Nus$, due to the disappearance of the corner flow for increasing $\ell$.

\vspace{-7 mm}

%


\end{document}